\documentclass[prd,aps,nofootinbib,notitlepage,showpacs,showkeys,preprintnumbers]{revtex4-1}
\usepackage{graphicx,epsf,amsmath,amsfonts,amssymb,amsbsy}
\usepackage{epsfig}
\usepackage{epstopdf}
\usepackage{tikz}
\usepackage{verbatim}

\newcommand{\ds}{\displaystyle}
\newcommand{\vev}[1]{\langle#1\rangle}
\newcommand{\mat}{\left ( \begin{array}}
\newcommand{\emat}{\end{array} \right )}
\newcommand{\vect}{\left ( \begin{array}{c}}
\newcommand{\evect}{\end{array} \right )}

\newcommand{\Det}{\mathop{\rm Det}\nolimits}

\begin{document}

\title{ \bf Charged pion condensation in dense quark matter: Nambu--Jona-Lasinio model study}
\author{T. G. Khunjua$^{1,2}$,  K. G. Klimenko$^{3}$, R. N. Zhokhov$^{3,4}$}
\vspace{1cm}
\affiliation{$^{1)}$ Faculty of Physics, Moscow State University,
119991, Moscow, Russia}
\affiliation{$^{2)}$ University of Georgia, Tbilisi, Georgia}
 \affiliation{$^{3)}$ Logunov Institute for High Energy Physics,
NRC "Kurchatov Institute", 142281, Protvino, Moscow Region, Russia}
\affiliation{$^{4)}$  Pushkov Institute of Terrestrial Magnetism, Ionosphere and Radiowave Propagation (IZMIRAN),
108840 Troitsk, Moscow, Russia}

\begin{abstract}
In this short review we tried to give an outline of investigations of charged pion condensation (PC) in dense baryonic (quark) 
matter in the framework of effective Nambu--Jona-Lasinio (NJL) type models. The possibility of charged PC phase in dense quark 
matter with isospin asymmetry is investigated. First, it is demonstrated that this phase can be realized in the framework of 
massless NJL model. But the existence of this phase is enormously fragile to the values of current quark mass 
and we show that charged PC phase is forbidden in electrically neutral dense quark matter with $\beta$-equilibrium when 
current quark masses are close to their physical value of 5.5 MeV. Nevertheless, then it is shown that 
in real physical systems there could be conditions promoting the appearance of charged PC phenomenon in dense quark matter, 
namely, it was shown that if one includes into consideration the fact that system can have finite size,
then a dense charged PC phase can be realized there.  It was also demonstrated that the possibility of inhomogeneous pion
condensate might allow this phase to appear. And more recently it was revealed that there is another interesting factor that 
can induce a charged PC phase in dense quark matter even without isospin imbalance. It is a chiral imbalance of the system 
(non-zero difference between densities of left- and right-handed quarks). This results can be interesting in heavy ion 
collision experiments, where it is expected to get high baryon densities. It is of interest also in the context of the neutron 
stars, where quark matter might be realized in the core and very high baryon and isospin densities are attained.
\end{abstract}

\keywords{
NJL model; charged pion condensation; baryon density; isospin imbalance, chiral imbalance
}
\maketitle


\section{Introduction}
Recently, great efforts have been made in trying to understand the properties and the phase diagram of strongly interacting 
(quark) matter under high 
temperatures and/or baryon densities (or high values of the baryon chemical potential $\mu_B$). It is believed that quark matter 
under such extreme conditions can exist in cores of neutron stars, where baryon density is much higher than the density 
$\rho_0$ of 
ordinary nuclear matter, where $\rho_0$ is 0.15 baryon per fm$^{-3}$, or can be formed as a result of a collision of heavy ions \cite{Shuryak:1980tp}. 
(In this case there are no reasons to speak of protons and neutrons as particles that make up baryonic matter, and 
it would be more correct to say that we are dealing with dense quark matter.) However, in reality, within the framework of 
quantum chromodynamics (QCD), the fundamental theory designed to describe strongly interacting systems, it is rather 
problematic to get at least some information about the phase structure of dense quark matter. The fact is
that despite asymptotic freedom and 
feasibility of perturbative calculations at extremely high temperature and the baryon density in real physical conditions (in compact stars or heavy ion collision experiments) the temperature and the baryon density of 
quark matter are usually not that great, which leads to the fact that the effective interaction constant of the particles is quite large. 
Therefore, one of the main research methods in the framework of QCD, perturbative analysis over a small coupling constant, is not
applicable. Another well-known first-principle calculation approach to QCD is the method of lattice QCD simulations. But, 
unfortunately, due to the sign problem (complex fermion determinant) it encounters up to now insurmountable difficulties at 
$\mu_B\ne 0$.

Due to these reasons, in order to study the QCD phase diagram at nonzero baryon density,
effective field theories are employed and the most frequently used one is the Nambu–Jona-Lasinio (NJL) model \cite{njl} 
(for a review one can see Refs. \cite{Hatsuda:1994pi,Buballa:2003qv}). This model uses quarks as the degrees of freedom instead 
of hadrons, and it can be used as an effective field theory for QCD in the low energy regime in order to describe dense quark matter.
The most appealing feature of NJL model is the description of the phenomenon of
the dynamical chiral symmetry breaking when quarks get a rather large mass compared to the value of the current quark mass. 
The NJL model can be used to get insights not only in meson physics \cite{ebert}, but also in color superconductivity 
phenomenon \cite{alford,shovkovy}, etc.

As it was mentioned above, dense quark matter can be formed inside neutron stars, i.e. it consists of different numbers of 
$u$ and $d$ quarks (it is the so-called isotopic asymmetry of quark matter). Hence, the ground state of such matter is characterized
by baryon and isotopic (or isospin) densities and can be investigated in the framework of QCD or other alternative effective 
approaches using quark degrees of freedom and extended by baryon $\mu_B$ and isotopic (or isospin) $\mu_I$ chemical potentials. 
Then, as it was shown in Refs. \cite{Ebert:2005cs, Ebert:2005wr, He:2005tf, He:2005nk, He:2005sp, Barducci:2004tt,Loewe:2005yn, Splittorff:2000mm, Son:2000xc,Andersen:2007qv,KogutSinclair, Gupta:2002kp}, in the corresponding systems there might appear a 
Bose-Einstein condensate 
(even at $\mu_B=0$) of quark-antiquark pairs of the form $\bar u\gamma^5 d$, which has quantum numbers of charged pions. 
In this case the charged pion condensation (PC) phase is realized in matter. However, the presence of this phase of dense quark 
matter is not a well-established fact up to now. Indeed, it was shown in Refs. \cite{Ebert:2005cs, Ebert:2005wr} that charged PC phase 
with nonzero quark density occupies a very narrow part of the phase diagram of the massless NJL model when $\mu_B\ne 0$ 
and $\mu_I\ne 0$. But this phase is extremely fragile to the effect of explicit chiral symmetry breaking 
via a non-zero current quark mass and is ruled out from the phase diagram of electrically neutral dense quark matter with 
$\beta$-equilibrium, when quark masses are larger than or of 
the order of 10 keV (see, e.g., in Refs. \cite{Abuki:2008wm, Anglani:2010ye, Abuki:2008tx, Andersen:2007qv, Abuki:2008nm}). 
The same results were found in the framework of the Polyakov NJL model  \cite{Abuki:2008tx}, where it was shown that  
the value of isospin chemical potential $\mu_I$ turns out to be not large enough for charged PC phase to takes place in electrically
neutral dense quark matter. Similar qualitative picture of the charged PC has been also found in an analysis in 
the framework of the toy (1+1)-dimensional massive Gross-Neveu model with isospin and baryon chemical potentials 
\cite{Ebert:2009ty}. In particular, for example, it was found
that 
baryon density is equal to zero in the charged PC phase of this model, i.e. this model does not predict 
dense quark matter with charged PC phenomenon.

Before continuing our discussion on the fate of the charged PC phase in dense quark matter, let us say a few words about the 
PC phenomenon in dense nuclear matter. It is clear that in this case the baryon density is around or greater than 
the normal baryon density $\rho_0$, i.e. much less than in dense quark matter. It is clear that the main elementary objects 
forming nuclear matter are nucleons and pions. Therefore, a long time ago, in the early seventies, it was understood that pions 
and pion condensates can play a significant 
role in the physical processes of dense nuclear matter \cite{Migdal} (see also, for example, the reviews \cite{Migdal2,Yakovlev}).
As for physics of neutron stars, where in nuclear matter there is a large isotopic asymmetry (the predominance of neutrons 
over protons), here, of course, processes with the formation of charged pion condensate can play a prominent role. Indeed,
in a series of papers \cite{Tatsumi,Takatsuka,Suzuki} it was shown that charged PC phase has a significant effect on the cooling 
scenario of a neutron star, on its equation of state, on the phenomenon of superfluidity of nucleons in neutron stars, etc. 
The main thing for us is that different effective approaches (such as the $\sigma$-model, etc.) used in these studies to describe 
neutron matter, in principle, do not reject the possibility of the existence of a phase with charged PC. 
(Though let us note that the $s$-wave charged PC is considered highly unlikely to be realized
 in 
dense matter \cite{Ohnishi:2008ng} but it was argued that $p$-wave charged PC is possible \cite{EricsonWeise,Wakasa:1997zz}.)
In contrast, as it follows from the above discussion, the possibility of the charged PC phenomenon in dense quark matter, in which
baryon density is much higher than $\rho_0$, described in the framework of the effective NJL model remains in question.

In this mini-review we discuss the present status of the charged PC phase in dense quark matter and, especially, pay attention
to the factors promoting the emergence of charged PC phenomenon in dense quark matter. 
Namely, in the paper \cite{Ebert:2011tt} it was found that this phase might be realized in dense baryonic system that has 
finite size. It was also demonstrated in \cite{Gubina:2012wp} that in dense baryonic matter spatially inhomogeneous 
charged PC might appear, and more recently it was found in 
\cite{Khunjua:2017khh,  Khunjua:2018sro, Khunjua:2017mkc, Ebert:2016hkd, Khunjua:2018jmn} that chiral imbalance is
another interesting factor that can induce a charged PC phase. Let us recall that chiral imbalance is a 
nonzero difference
between densities of left-handed and right-handed fermions and it may appear in the fireball
after heavy ion collisions and possibly lead to the so-called chiral magnetic effect \cite{Kharzeev:2007jp, Fukushima:2008xe}. 
It is a remarkable phenomenon that stems from highly nontrivial interplay of chiral symmetry, chiral (axial) anomaly of QCD and the
topological structure of gluon configurations. It might be also realized in compact stars \cite{Khunjua:2018jmn} or even 
in condensed matter systems \cite{Li:2014bha, Braguta:2019pxt}.

The promotion of the charged PC phase in dense quark matter by finite size and spatial inhomogeneity of the charged pion 
condensate was demonstrated using a (1+1)-dimensional toy model with four-quark interactions, whereas the generation 
of the charged PC phase by chiral imbalance was shown as in (1+1)-dimensional toy model 
\cite{Khunjua:2017khh, Ebert:2016hkd} as well as in more realistic effective QCD motivated NJL model \cite{Khunjua:2017mkc, Khunjua:2018sro}.

The paper is organized as follows. In Sec. II NJL model
with two quark flavors ($u$ and $d$ quarks) and with two kinds of chemical potentials, $\mu_B$ and $\mu_I$, is presented and the 
charged PC phenomenon in dense isospin asymmetric quark matter is discussed. Here we show that it is unlikely to 
appear in the simplest case. Then in Sec. III it is revealed that there exist the conditions and factors promoting the 
appearance of charged PC in dense quark matter and different possibilities are discussed. In Sec. III A  
the influence of finite size effect and nontrivial space topology is considered, then in Sec. III B the possibility of 
inhomogeneous charged PC is discussed, finally in Sec. III C it is shown that charged PC can 
appear in dense chiral imbalanced quark matter. Sec. IV contains the summary and conclusions. Throughout the paper, for simplicity,
all the considerations are made in the case of zero temperature $T=0$.

\section{Charged pion condensation in the framework of NJL model}

\subsection{The model and its thermodynamic potential}

Since the considered studies are mostly performed in the framework of the (3+1)-dimensional NJL model, let us recall  
the two flavored NJL model
Lagrangian (there are no gluons and it is symmetric under global color $SU(N_c)$ group). It has the following form
\begin{eqnarray}
&&  L_{NJL}=\bar q\Big [\gamma^\nu\mathrm{i}\partial_\nu-m
\Big ]q+ \frac
{G}{N_c}\Big [(\bar qq)^2+(\bar q\mathrm{i}\gamma^5\vec\tau q)^2 \Big
],  \label{1Lag}
\end{eqnarray}
where $q=(q_u,q_d)^T$ is a flavor doublet and $q_u$ and $q_d$ ($u$ and $d$ quarks) are four-component Dirac spinors as well as color $N_c$-plets. The summation over flavor, color, and spinor indices is implied; $\tau_k$ ($k=1,2,3$) are Pauli matrices. And $m$ is current quark mass.

Moreover, since in (3+1)-spacetime dimensions the four-fermion interaction operator has a mass dimension 6, 
the NJL model is not renormalizable and may be considered only as an effective QCD-like model. 
Hence, some regularization is needed in order to
avoid usual ultraviolet divergences and throughout the paper we will use the simplest momentum cutoff regularization, i.e.
the integration region in all momentum integrals is restricted by a cutoff $\Lambda$,  $|\vec p|<
\Lambda$. And the model parameters are fixed as follows:
$G =15.03$ GeV$^{-2}$, $\Lambda =0.65$ GeV. 

Since our paper is devoted to the dense baryonic (quark) matter at various conditions,  now, as a starting point,
let us first consider the dense quark matter with isospin asymmetry. The Lagrangian of two-flavored NJL model with baryon 
and isopin chemical potentials has the following form
\begin{eqnarray}
&& L= L_{NJL}+\bar q\Big[
\frac{\mu_B}{3}\gamma^0+\frac{\mu_I}2\gamma^0\tau_{3}
\Big ]q.
\label{1mu}
\end{eqnarray}
Since the Lagrangian (2) contains baryon $\mu_B$ and isospin $\mu_I$ chemical potentials, it can describe quark matter 
with nonzero baryon $n_B$ and isospin $n_I$ densities. Baryon $n_B$ and isospin $n_I$ densities are 
the quantities, thermodynamically conjugated to chemical potentials $\mu_B$ and $\mu_I$, respectively, and can be obtained 
if one takes a derivative of the thermodynamic potential (TDP) of the system (\ref{1mu}) with respect to $\mu_B$ and $\mu_I$ 
chemical potentials.

The quantities $n_B$ and $n_I$ are densities of conserved charges corresponding to the invariance of Lagrangian (\ref{1mu}) 
with respect to the abelian groups $U_B(1)$, $U_{I_3}(1)$, where \footnote{\label{f1,1}
Recall that the following relations hold~~
$\exp (\mathrm{i}\alpha\tau_3)=\cos\alpha
+\mathrm{i}\tau_3\sin\alpha$.
}
\begin{eqnarray}
U_B(1):~q\to\exp (\mathrm{i}\alpha/3) q;~\;\;\;\;\;\;\;\;\;\;\;\;\;\;\;\;\;\;
U_{I_3}(1):~q\to\exp (\mathrm{i}\alpha\tau_3/2) q.
\label{20011}
\end{eqnarray}
So one can show that $n_B=\bar q\gamma^0q/3$ and $n_I=\bar q\gamma^0\tau^3 q/2$. One can also see that, in addition to 
(\ref{20011}), Lagrangian (\ref{1mu}) is invariant under the transformation of 
the electromagnetic $U_Q(1)$ group,
\begin{eqnarray}
U_Q(1):~q\to\exp (\mathrm{i}Q\alpha) q,
\label{2002}
\end{eqnarray}
where $Q={\rm diag}(2/3,-1/3)$.

In order to find the thermodynamic potential of the system it is convenient to use a semibosonized version of the 
Lagrangian (\ref{1mu}), with auxiliary composite bosonic fields $\sigma (x)$ and $\pi_a (x)$ $(a=1,2,3)$,
\begin{eqnarray}
\widetilde L\ds &=&\bar q\Big [\gamma^\rho\mathrm{i}\partial_\rho-m
+\mu\gamma^0
+ \nu\tau_3\gamma^0
-\sigma
-\mathrm{i}\gamma^5\pi_a\tau_a\Big ]q
 -\frac{N_c}{4G}\Big [\sigma\sigma+\pi_a\pi_a\Big ].
\label{L2}
\end{eqnarray}
Here the summation over repeated indices is implied.
Hereafter, we will use the notations $\mu\equiv\mu_B/3$ for quark chemical potential and $\nu\equiv\mu_I/2$.
One can get the equations of motion for the 
bosonic fields from the auxiliary Lagrangian (\ref{L2}),
\begin{eqnarray}
\sigma(x)=-2\frac G{N_c}(\bar qq);~~~\pi_a (x)=-2\frac G{N_c}(\bar q
\mathrm{i}\gamma^5\tau_a q).
\label{200}
\end{eqnarray}
 Utilizing the equations (\ref{200}), it is possible to show that the semibosonized Lagrangian $\widetilde L$ (\ref{L2}) is equivalent to the initial Lagrangian (\ref{
 1mu}). Note that the auxiliary composite bosonic field $\pi_3 (x)$ can be identified with the physical $\pi^0(x)$-meson field, whereas the physical $\pi^\pm (x)$-meson fields are the following combinations of the composite bosonic fields, $\pi^\pm (x)=(\pi_1 (x)\mp i\pi_2 (x))/\sqrt{2}$.
In addition, let us note that the composite bosonic fields (\ref{200}) are transformed under the isospin $U_{I_3}(1)$ group in the following manner (it can be shown from (\ref{200}) and footnote \ref{f1,1}):
\begin{eqnarray}
U_{I_3}(1):~&&\sigma\to\sigma;~~\pi_3\to\pi_3;~~\pi_1\to\cos
(\alpha)\pi_1+\sin (\alpha)\pi_2;~~\pi_2\to\cos (\alpha)\pi_2-\sin
(\alpha)\pi_1.
\label{201}
\end{eqnarray}
From the auxiliary Lagrangian (\ref{L2}), it is possible to obtain in the leading order of the large-$N_c$ expansion 
(i.e. in the one loop approximation) the effective action ${\cal S}_{\rm {eff}}(\sigma,\pi_a)$ of the model. This quantity is a 
functional of the auxiliary bosonic $\sigma (x)$ and $\pi_a (x)$ fields and has the following form:
$$
\exp(\mathrm{i}{\cal S}_{\rm {eff}}(\sigma,\pi_a))=
  N'\int[d\bar q][dq]\exp\Bigl(\mathrm{i}\int\widetilde L\,d^4 x\Bigr).
$$
It is easy to obtain for ${\cal S}_{\rm {eff}}(\sigma,\pi_a)$ the following expression 
\begin{equation}
{\cal S}_{\rm {eff}}
(\sigma(x),\pi_a(x))
=-N_c\int d^4x\left [\frac{\sigma^2+\pi^2_a}{4G}
\right ]+\tilde {\cal S}_{\rm {eff}},
\label{L3}
\end{equation}
in which the last term (i.e. the term $\tilde {\cal S}_{\rm {eff}}$ in (\ref{L3})) is a quark contribution to the 
effective action,
\begin{eqnarray}
\exp(\mathrm{i}\tilde {\cal S}_{\rm {eff}})&=&N'\int [d\bar
q][dq]\exp\Bigl(\mathrm{i}\int\Big\{\bar q\big
[\gamma^\rho\mathrm{i}\partial_\rho-m +\mu\gamma^0+
\nu\tau_3\gamma^0
-\sigma -\mathrm{i}\gamma^5\pi_a\tau_a\big
]q\Big\}d^4 x\Bigr)\nonumber\\
&=&[\Det D]^{N_c},
 \label{4}
\end{eqnarray}
where $N'$ is a normalization constant and 
$
D\equiv\gamma^\nu\mathrm{i}\partial_\nu-m +\mu\gamma^0
+ \nu\tau_3\gamma^0-\sigma (x) -\mathrm{i}\gamma^5\pi_a(x)\tau_a
$
is the Dirac operator acting in the flavor, spinor and coordinate spaces. 
Now let us use the general formula $\Det D=\exp {\rm Tr}\ln D$ and obtain the following expression for the effective action (\ref{L3})
\begin{equation}
{\cal S}_{\rm {eff}}(\sigma(x),\pi_a(x))
=-N_c\int
d^4x\left[\frac{\sigma^2(x)+\pi^2_a(x)}{4G}\right]-\mathrm{i}N_c{\rm
Tr}_{sfx}\ln D,
\label{6}
\end{equation}
where the Tr operation is performed in spinor ($s$), flavor
($f$) and four-dimensional space-time coordinate ($x$) spaces, respectively.

Now the ground state expectation values $\vev{\sigma(x)}$ and
$\vev{\pi_a(x)}$ of composite bosonic fields can be found from 
the saddle point equations,
\begin{eqnarray}
\frac{\delta {\cal S}_{\rm {eff}}}{\delta\sigma (x)}=0,~~~~~
\frac{\delta {\cal S}_{\rm {eff}}}{\delta\pi_a (x)}=0,
\label{05}
\end{eqnarray}
where $a=1,2,3$. The knowledge of values $\vev{\sigma(x)}$ and
$\vev{\pi_a(x)}$, namely their behaviour with respect to chemical potentials, provides us with phase structure of the model. 

It is clear that if there is a non-zero ground state expectation values of composite bosonic $\sigma(x)$ and/or $\pi_3(x)$ 
fields, i. e. $\vev{\sigma(x)}\ne 0$ and/or $\vev{\pi_3(x)}\ne 0$,  then in the chiral limit we have a spontaneous breaking of 
the axial isospin $U_{AI_3}(1)$ symmetry in the model (if we consider the chiral limit $m=0$, when the Lagrangian (\ref{1mu}) 
is symmetric in addition with respect to $U_{AI_3}(1)$ group, which is defined as 
$U_{AI_3}(1):~q\to\exp (\mathrm{i}\alpha\gamma^5\tau_3/2) q$), whereas if other component of pion field acquire non-zero ground 
state expectation value, i. e. $\vev{\pi_1(x)}\ne 0$ and/or $\vev{\pi_2(x)}\ne 0$, then, as it is obvious from (\ref{201}), 
the isospin symmetry ($U_{I_3}(1)$ group) is spontaneously broken down. Since in this latter case the condensates, or the ground 
state expectation values, of the charged pion fields $\pi^+(x)$ and $\pi^-(x)$ are non-zero, this phase is usually called the 
charged PC phase. Furthermore, one can see from  (\ref{200}) that the nonzero condensates $\vev{\pi_{1,2}(x)}$ (or $\vev{\pi^\pm(x)}$) are not invariant with respect to the transformations of the electromagnetic group $U_Q(1)$ (\ref{2002}) and this leads to spontaneous breaking of electromagnetic $U_Q(1)$ symmetry. Hence one can say that the electromagnetic invariance in the charged PC phase is broken down and superconductivity is an unavoidable property of the charged PC phase.

As a rule one supposes that in the ground state of the system expectation values $\vev{\sigma(x)}$ and $\vev{\pi_a(x)}$ do not depend on space-time coordinates $x$,
\begin{eqnarray}
\vev{\sigma(x)}\equiv M-m,~~~\vev{\pi_a(x)}\equiv \pi_a, \label{8}
\end{eqnarray}
where $M$ and $\pi_a$ ($a=1,2,3$) are already constant quantities and they can be found as coordinates of the global minimum point of the
thermodynamic potential $\Omega (M,\pi_a)$.

The TDP is defined by the following expression in the leading order of the large-$N_c$ expansion:
\begin{equation}
\int d^4x \Omega (M,\pi_a)=-\frac{1}{N_c}{\cal S}_{\rm
{eff}}\big (\sigma(x),\pi_a (x)\big )\Big|_{\sigma
(x)=M-m, \pi_a(x)=\pi_a}.\label{08}
\end{equation}
To find the phase portrait of the model one needs to investigate the dependence on $\mu,\nu,$ of the global minimum point 
(GMP) of the TDP $\Omega (\sigma,\pi_a)$ vs $\sigma,\pi_a$. To this end let us simplify the task and note that due to a
$U_{I_3}(1)$ invariance of the model, the TDP (\ref{08}) depends really only on the three combinations of bosonic fields, 
$\sigma$, $\pi_3$ and $\pi_1^2+\pi_2^2$ (this can be seen from (\ref{201})). So without loss of generality, it can be 
assumed that $\pi_2=0$. Moreover, it is also possible to prove that $\pi_3=0$ at $m\ne 0$ (see, e.g., in \cite{Fedotov}).
Hence, one can study the TDP as a function of only two variables,
$M\equiv\sigma+m$ and $\Delta\equiv\pi_1$. It means that we use the following ansatz
\begin{eqnarray}
\vev{\sigma(x)}=M-m,~~~\vev{\pi_1(x)}=\Delta,~~~\vev{\pi_2(x)}=0,~~~ \vev{\pi_3(x)}=0. \label{06}
\end{eqnarray}

The expression for the
TDP (\ref{08}) of the dense (baryon density) and isospin asymmetric quark matter can be easily obtained in the mean field 
approximation and it has the form 
\begin{eqnarray}
\Omega(M,\Delta)=\frac{(M-m)^2+\Delta^2}{4G}-2\int\frac{d^3p}{(2\pi)^3}\Big\{E_
\Delta^-+E_\Delta^++(\mu-E_\Delta^-)\theta(\mu-E_\Delta^-)+
(\mu-E_\Delta^+)\theta(\mu-E_\Delta^+) \Big\},
\label{2omega}
\end{eqnarray}
where there have been introduced the following notations $E_\Delta^\pm=\sqrt{(E^\pm)^2+\Delta^2}$,
and $E^\pm=E\pm\nu$, $E=\sqrt{\vec p^2+M^2}$.
Let us recall also that the integration in the momentum space is restricted by a cutoff 
$|\vec p|<\Lambda$. From (\ref{2omega}) it is possible to obtain the gap equations
\begin{eqnarray}
0=\frac{\partial\Omega (M,\Delta)}{\partial M}&\equiv&
\frac{M-m}{2G}-2M\int\frac{d^3p}{(2\pi)^3E}\Big\{\frac{\theta(E_
\Delta^+-\mu)E^+}{E_\Delta^+}+
\frac{\theta(E_\Delta^--\mu)E^-}{E_\Delta^-} \Big\},\nonumber\\
0=\frac{\partial\Omega (M,\Delta)}{\partial\Delta}&\equiv&
\frac{\Delta}{2G}-2\Delta\int\frac{d^3p}{(2\pi)^3}\Big\{\frac{\theta(
E_\Delta^+-\mu)}{E_\Delta^+}+
\frac{\theta(E_\Delta^--\mu)}{E_\Delta^-} \Big\}.
\label{4gaps}
\end{eqnarray}
Note that the quark gap $M$ is just the dynamical quark mass.

 \subsection{Phase diagram at $\nu \ne 0$ and $\mu\ne 0$ in the chiral limit}

First, let us discuss the phase diagram and the charged PC in dense quark matter with isospin asymmetry in the 
chiral limit. This case has been considered and the phase portrait of the model at both $\mu_I \ne 0$ and $\mu\ne 0$ has been 
obtained in \cite{Ebert:2005cs, Khunjua:2017mkc}. Since, in general, it is a rather hard task to consider the phase diagram at 
the physical point, in this paper, for simplicity, the current quark mass $m$ was assumed to be zero, i. e. the chiral limit was used.
The corresponding $(\nu,\mu)$-phase portrait is depicted in Figure 1. Let us discuss its content. One can see there is a huge region of
PC phase at this phase diagram. Note that it has been shown already in \cite{barducci, 
 He:2005tf, He:2005nk, He:2005sp} that at $m\ne 0$ the charged PC phase
 appears at $\mu_I>m_\pi$. But in the chiral limit (see in Figure 1) the same 
 relation $\mu_I>m_\pi$ is also valid since $m_\pi=0$ in this case. 
 
 \begin{figure*}
 \centering
\includegraphics[width=0.55\textwidth]{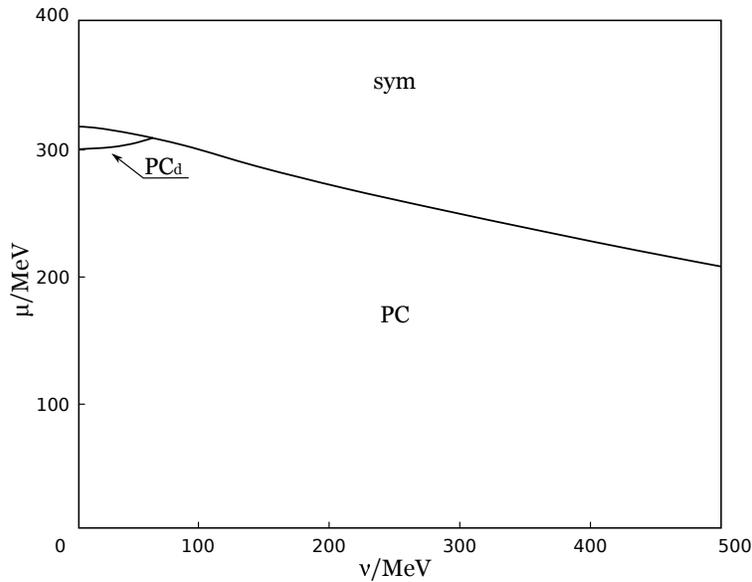}
\caption{The ($\nu$, $\mu$) phase portrait in the chiral limit in the framework of NJL model. 
 PC phase is charged pion condensation phase.
Subscript $d$ means that in this phase baryon
density is nonzero. The sym denotes the symmetric phase. 
}
\end{figure*}

 Now let us discuss all possible phases in this case.
The system of the gap equations (\ref{4gaps}) at $m=0$ has three different types of solutions: {\bf
(i)} $M=0$, $\Delta =0$, {\bf (ii)} $M\ne 0$, $\Delta =0$, {\bf (iii)}
$M=0$, $\Delta\ne 0$. The {\bf
(i)} solution corresponds to the symmetric phase, where all condensates are zero.
If the GMP of the TDP is located at the point of the form {\bf (ii)}, 
then the charged pion condensate is equal to zero in the ground state and the
quark condensate is nonzero. This phase is called a chiral symmetry breaking (CSB) phase. 
Finally, the {\bf (iii)}  solution corresponds to charged PC phase, where the quark condensate is equal to zero, 
but the charged pion one has a nonzero value.

If the GMP of the TDP is achieved on the solution of type {\bf (i)}, then one can see that the
ground state of the system is invariant with respect to the initial symmetry
group of the model, which at $m=0$ (in the chiral limit) and $\mu_I \ne 0$ is
$\rm U_B(1)\times U_{I_3L}(1)\times U_{I_3R}(1)$. The {\bf (ii)}-type solution (CSB phase) corresponds to
the phase symmetric with respect to $\rm U_B(1)\times U_{I_3}(1)$ group.
In the charged PC phase the ground state is invariant under transformations from
$\rm U_B(1)\times U_{AI_3}(1)$ group. Moreover, it is clear that in the ground state of quark matter with non-zero 
charged pion condensate parity is also broken down. 

Let us return to the discussion of the ($\nu$, $\mu$)-phase portrait of the model (see in Figure 1).
One can see that apart from the PC phase there is a region of the PC$_{d}$ phase at the phase diagram of Figure 1. This phase denotes the ground state
of quark matter, in which besides non-zero charged pion condensate the baryon density is also non-zero. 
This consideration in terms of NJL model shows that the charged PC can be realized in dense quark matter with isospin asymmetry. But let us recall once again that considerations have been performed in the chiral limit when pions are a genuine Goldstone bosons and have zero masses.

 \subsection{Account of electric neutrality and $\beta$-equilibrium conditions in the chiral limit}

After the prediction of the charged PC phenomenon in the dense and isospin asymmetric quark matter (that can be realized, for example, in relativistic heavy ion collisions or inside the cores of the compact stars) it was interesting to 
take into account the charge neutrality and $\beta$-equilibrium constraints 
that are important inside compact stars. The possibility of the charged PC phenomenon 
in cold and electrically neutral dense baryonic matter
was investigated in $\beta$-equilibrium in \cite{Ebert:2005wr}, where the
consideration was also performed in the chiral limit.

Let us first elaborate on how to implement all these constraints.
As it has been said above, the electric charge is conserved in the model and it can be written as $Q={\rm diag}(2/3,-1/3)=I_3+B/2$, where $I_3=\tau_3/2$ is the
generator of the third isospin component in the flavor space and $B={\rm diag}(1/3,1/3)$
is the baryon charge generator. 
Due to the $\beta$-equilibrium requirement, one should incorporate
electrons in our consideration. They are necessary to neutralize the electric charge of quarks. Moreover, 
$\beta$-equilibrium of quark matter means that all $\beta$-processes that include quarks and leptons should go in both 
directions with equal rates. 
As it was argued above, full Lagrangian of the system should include the electron field and electron chemical potential, 
but since our system is in $\beta$-equilibrium, 
one can show that the electron chemical potential should be equal to 
charge chemical potential of quarks with opposite sign.

It was also supposed that one can neglect the mass of the electron and put it to zero for simplicity for their mass is 
quite small in the considered scale.

Since in the system composed of quarks and electrons both baryon charge and electric charge are conserved quantities, its properties 
in equilibrium are described by the following Lagrangian
\begin{eqnarray}
L=
L_{NJL}+\bar e\gamma^\nu i\partial_\nu e+\mu_BN_B+\mu_QN_Q,
  \label{3}
\end{eqnarray}
where $L_{NJL}$ is the  Lagrangian of NJL model (\ref{1Lag}), $e$ is the electron spinor field; $N_B$ and $N_Q$ are baryon and electric charge
density operators, respectively, and $\mu_B$, $\mu_Q$ are corresponding chemical potentials. 
It is clear that
\begin{eqnarray}
N_B=\bar q B\gamma^0q,~~~~~N_Q=\bar q Q\gamma^0q-\bar e\gamma^0e.
\label{4}
\end{eqnarray}
Hence the Lagrangian (\ref{3}) can be presented in the form
\begin{eqnarray}
L=
L_{NJL}+\bar e\gamma^\nu i\partial_\nu e+(\mu_B/3+\mu_Q/6)\bar q \gamma^0q+
\mu_Q\bar qI_3\gamma^0q-\mu_Q\bar e\gamma^0e.    
  \label{3*}
\end{eqnarray}
In particular, it is clear from this relation that $\mu_Q$ is just the isospin chemical potential $\mu_I$.

The TDP in this case 
was obtained in \cite{Ebert:2005wr} in the mean field approximation. In contrary to the previous sections where the large $N_{c}$ approximation was used, here (since in neutral quark matter the electrons are taken into account and the count of degrees of freedom matters) it is important that there are three colors of quarks ($N_{c}=3$) and one should work in the mean field approximation. One should be careful as in this paper slightly different definitions are used compared to \cite{Ebert:2005wr}.  In the chiral limit the TDP
has the following form (the definition of the TDP is the same as in (\ref{08}))
\begin{eqnarray}
\Omega(M,\Delta)=-\frac{\mu_Q^4}{36\pi^2}+\frac{M^2+\Delta^2}{4G}
-\sum_a\int\frac{d^3p}
{(2\pi)^3}~|E_a|,
\label{5Omega}
\end{eqnarray}
where the first term in the right hand side of (\ref{5Omega}) is the TDP of free
massless electrons. The summation in the 
third term of (\ref{5Omega}) runs over all
quasiparticles ($a=u,d,\bar u,\bar d$), where
\begin{eqnarray}
E_u=E_\Delta^--\bar\mu,~~~~~~&& E_{\bar u}=E_\Delta^++\bar\mu,
\nonumber\\
E_d=E_\Delta^+-\bar\mu,~~~~~~&& E_{\bar d}=E_\Delta^-+\bar\mu,
\label{6}
\end{eqnarray}
and $E_\Delta^\pm=\sqrt{(E^\pm)^2+\Delta^2}$, $E^\pm=E\pm\mu_Q/2$,
$E=\sqrt{\vec p^2+M^2}$, $\bar\mu=\mu_B/3+\mu_Q/6$. 
As always, in order to
avoid usual ultraviolet divergences, the integration region in
(\ref{5Omega}) is restricted by a cutoff $\Lambda$, i.e. $|\vec p|<
\Lambda$. 
The gap equations that can be obtained from (\ref{5Omega}) look like
\begin{eqnarray}
0=\frac{\partial\Omega (M,\Delta)}{\partial M}&\equiv&
\frac{M}{2G}-2M\int\frac{d^3p}{(2\pi)^3E}\Big\{\frac{
\theta(E_\Delta^+-\bar\mu)E^+}{E_\Delta^+}+
\frac{\theta(E_\Delta^--\bar\mu)E^-}{E_\Delta^-} \Big\},\nonumber\\
0=\frac{\partial\Omega (M,\Delta)}{\partial\Delta}&\equiv&
\frac{\Delta}{2G}-2\Delta\int\frac{d^3p}{(2\pi)^3}\Big\{\frac{\theta(
E_\Delta^+-\bar\mu)}{E_\Delta^+}+
\frac{\theta(E_\Delta^--\bar\mu)}{E_\Delta^-} \Big\}.
\label{8}
\end{eqnarray}
In order to impose the neutrality constraint locally, we should look for the ground state of the system with the electric charge density $n_Q\equiv
-N_{c}\partial\Omega /\partial\mu_Q=-3\partial\Omega /\partial\mu_Q$ equals identically to zero. One can rephrase it in
other words as the study of the GMP of the TDP $\Omega(M,\Delta)$
under the constraint $n_Q=0$ that has the following form
\begin{eqnarray}
0=n_Q\equiv\frac{\mu_Q^3}{3\pi^2}+\int\frac{d^3p}{(2\pi)^3}\Big\{
\theta(\bar\mu-E_\Delta^+)+\theta(\bar\mu-E_\Delta^-)+
3\theta(E_\Delta^+-\bar\mu)\frac{E^+}{E_\Delta^+}-
3\theta(E_\Delta^--\bar\mu)\frac{E^-}{E_\Delta^-} \Big\}.
\label{9}
\end{eqnarray}
Then, it was shown in \cite{Ebert:2005wr} by numerical investigations of the gap and neutrality equations (\ref{8}) and (\ref{9}) that
at sufficiently small quark chemical potential,  namely, at $\mu<\mu_{1c}\approx$
301 MeV, the ground state of the system corresponds to CSB phase. As it has been discussed above, in 
this case the GMP has the form {\bf (ii)} and the dynamical quark mass $M\approx$ 301 MeV. 
There is no need to neutralize the quark electric charge because in this
phase both the baryon and the isospin densities equals to zero. It means that $\mu_{I}=\mu_{Q}=0$ in this phase. 

\begin{figure}
  \centering
  \includegraphics[width=0.65\textwidth]{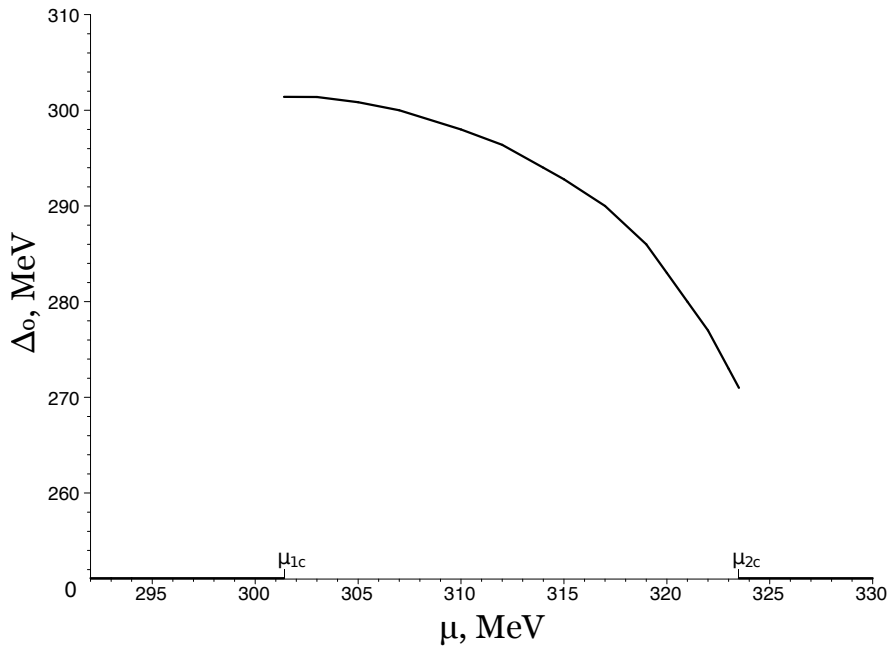}
  \caption{The value of charged pion condensate $\Delta_0$ as a function of quark number chemical potential $\mu$ 
  in electric charge neutral quark matter.}
\label{plot:2}
\end{figure}

At $\mu_{1c}<\mu<\mu_{2c}\approx$ 323.5 MeV the GMP has already the form {\bf (iii)}, meaning that in this case the charged 
PC phase is realized. In this phase 
the baryon density is non-zero so it is PC$_{d}$ phase. At Figure 2 the pion condensate $\Delta_0$ is depicted as a function of $\mu$. Then, at larger values of the chemical potential $\mu$ (i.e. at
$\mu_{2c}<\mu$ at Figure 2) symmetric phase is realized, where both condensates are zero. So one can see that in
electrically neutral dense baryonic matter in $\beta$-equilibrium there might appear charged PC phenomenon.
Let us stress once more that in \cite{Ebert:2005wr} this result was obtained in the chiral limit.

 \subsection{Phase diagram at $\mu_I \ne 0$ and $\mu\ne 0$ at the physical point ($m\ne 0$)}

Though current quark mass is small compared to the scales studied in the paper, it was shown in 
\cite{He:2005tf, He:2005nk, He:2005sp} that charged PC phenomenon at not so large values of isospin imbalance is 
strongly influenced by current quark mass values. As we have already said above, charged PC phenomenon takes place only at values
of $\mu_{I}>m_{\pi}$. And at physical point $m_{\pi}\neq 0$ 
it happens only at values of isospin  chemical potential larger than approximately 139 MeV. So non-zero current quark mass can be of importance for the 
investigations of the phase structure.

After the prediction of the charged PC phase of dense matter, it was necessary to 
consider this phenomenon at the physical point. It means that one needs to use the Lagrangian  (\ref{1Lag}) and (\ref{1mu}) with
non-zero physical value of current (bare) quark mass $m\approx 5.5$ MeV. The consideration at the physical point but without electric neutrality condition 
\begin{figure*}
 \centering
\includegraphics[width=0.55\textwidth]{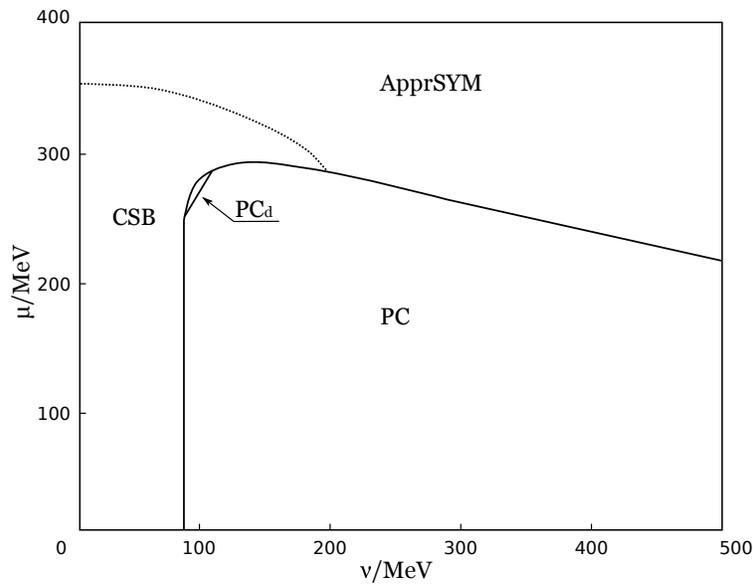}
\caption{The ($\nu$, $\mu$) phase portrait of the model at the physical point ($m\ne 0$). CSB phase denotes chiral symmetry breaking phase and
the notation 
ApprSYM means the phase where dynamical quark mass is close to the current quark mass value $M\approx m$, i.e. the chiral
symmetry is approximately restored in this phase. There is no a real phase transition between CSB and ApprSYM phases but a continuous crossover.  Other notations are the same as in Figure 1.}
\end{figure*}
showed that there remains marginal region of PC$_{d}$ phase at the phase portrait. The corresponding ($\nu,\mu$) phase diagram 
is shown at Figure 3 and one can see that in this case the region of PC$_{d}$ phase is much smaller than at Figure 1 corresponding to the chiral limit. Although, still charged PC phenomenon can take place in this case even at the physical point
(even though one should bear in mind that very small size of this phase and the fact that it is obtained in the framework of the effective 
model makes it not a very robust prediction). 

Now let us turn to the discussion on what happens if one includes into play the electric neutrality and beta-equilibrium conditions at the physical point.
In \cite{Abuki:2008wm, Anglani:2010ye, Andersen:2007qv} the consideration of the electrically neutral dense baryonic matter in $\beta$-equilibrium has been performed in the framework of NJL model at the physical point. 

In this case
the expression for the TDP (20) should be slightly changed, namely, the non-zero current quark mass should be added and one obtains
\begin{eqnarray}
\Omega(M,\Delta)=-\frac{\mu_Q^4}{36\pi^2}+\frac{(M-m)^2+\Delta^2}{4G}
-\sum_a\int\frac{d^3p}
{(2\pi)^3}~|E_a|.
\label{5}
\end{eqnarray}

Numerical studies of this case have been performed in \cite{Abuki:2008wm, Anglani:2010ye, Andersen:2007qv} and the phase structure has been obtained. In these studies the current quark mass $m$ has been treated as a free parameter, therefore, the corresponding pion mass $m_\pi$ at $\mu=T=0$ was a free parameter as well.

Let us say a couple of words on the results obtained.
At Figure 4 the $(\mu,\,m_\pi)$ phase diagram is presented.
The solid line denotes the border between the two phases, CSB phase (which is arranged at rather small values of $\mu$)
and ApprSYM one (this phase 
lies in  the region of large $\mu$).
ApprSYM means an approximate symmetrical 
phase and corresponds to a GMP of the TDP with $M\approx m\neq0$ and $\Delta=0$, but in contrast to the CSB phase, 
dynamical quark mass $M$ in ApprSYM drops rapidly and incessantly 
to the current quark mass value $m$ at increasing  temperature or baryon chemical  potential.
The bold dot at the end of the line is the critical endpoint of the first order phase transition. The shaded region 
denotes the charged PC phase.
Now if we go to the chiral limit $(m_{\pi}=0)$ (the line $\mu$ at $m_{\pi}=0$) these results are in good agreement with the ones discussed 
 above. There are two critical values of the chemical potential, $\mu_{c1}$
 and $\mu_{c2}$ that corresponding to the onset and disappearance of
 charged PC, respectively.
But increasing the value of the current quark mass the shaded
 region  shrinks until these two points meet, $\mu_{c1}\equiv \mu_{c2}$.
 They meet at the value of pion mass $m_{\pi}^{c}\sim 9\; \rm MeV$ that corresponds to a current quark mass of $m \sim 10\;\rm keV$, 
 much less than the physical values, $m \sim 5\;\rm MeV$. 
 The ($\mu$,$\mu_{Q}$) phase portrait at the physical point is 
 depicted at Figure 5 \cite{Abuki:2008wm}, where one can also see that the electric neutrality line does not cross the PC phase. 
Hence, one can infer that charged PC is an 
 extremely fragile phenomenon with respect to the explicit chiral symmetry breaking effect (non-zero current quark mass).
 
In the papers \cite{Abuki:2008wm,  Anglani:2010ye} it was also shown that the equality between the electric chemical 
potential (in essence the isospin one) and the pion mass in medium, $|\mu_{Q}|=m_{\pi}$, is a threshold to the charged PC phase provided the transition to the charged PC phase is of second order. 
And the absence of the charged PC can be understood in the following way. Note that increase of current 
quark mass leads to a drastic amplification of the vacuum pion mass because at small masses $m$ pion mass depends on quark 
mass as $m_\pi\propto\sqrt{m}$, whereas the change in the value of the quark mass in
 keV scale brings about no considerable modification to the value of electric chemical potential $\mu_Q$.
 
 \begin{figure*}
\includegraphics[width=0.49\textwidth]{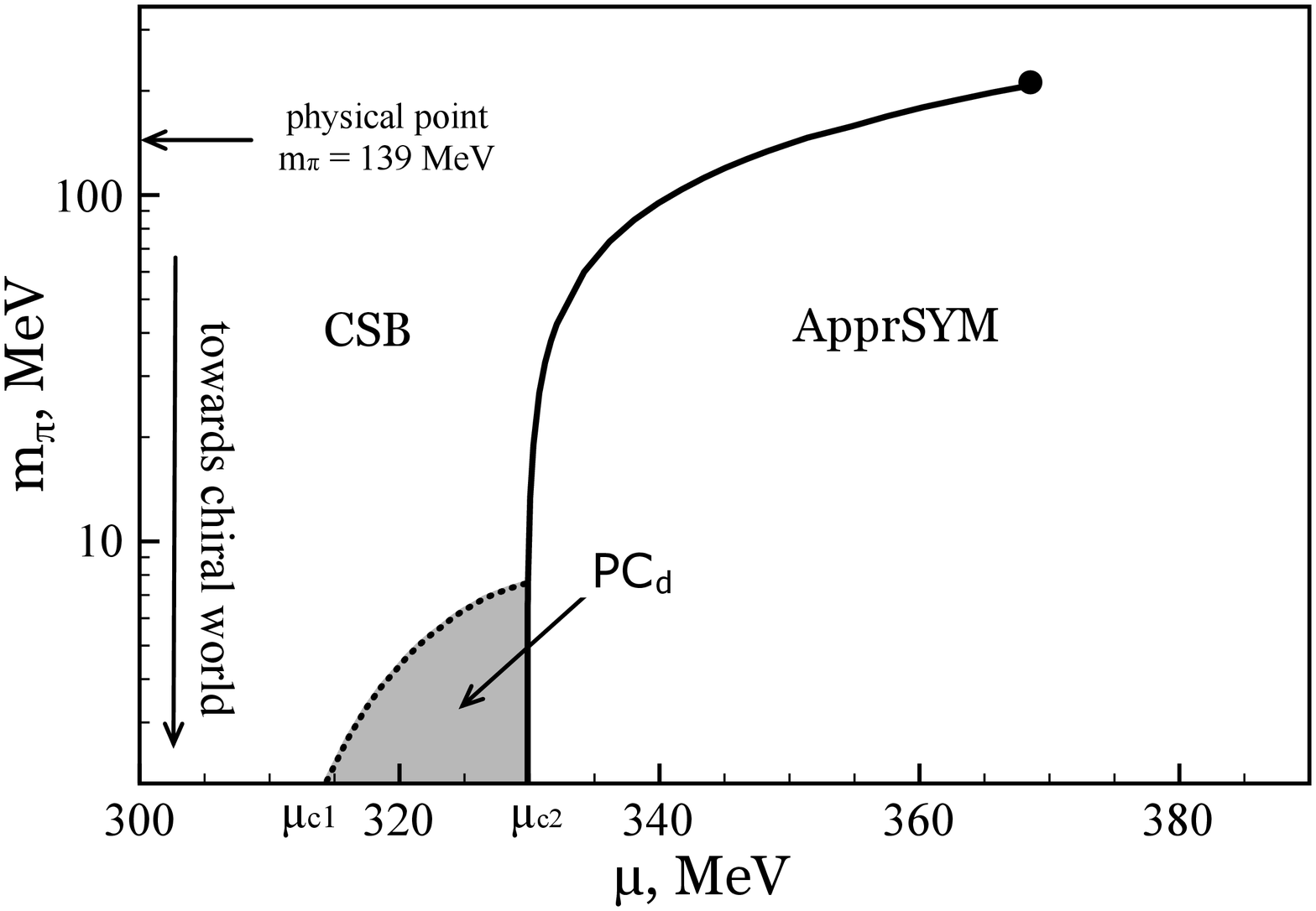}
\hfill
\includegraphics[width=0.49\textwidth]{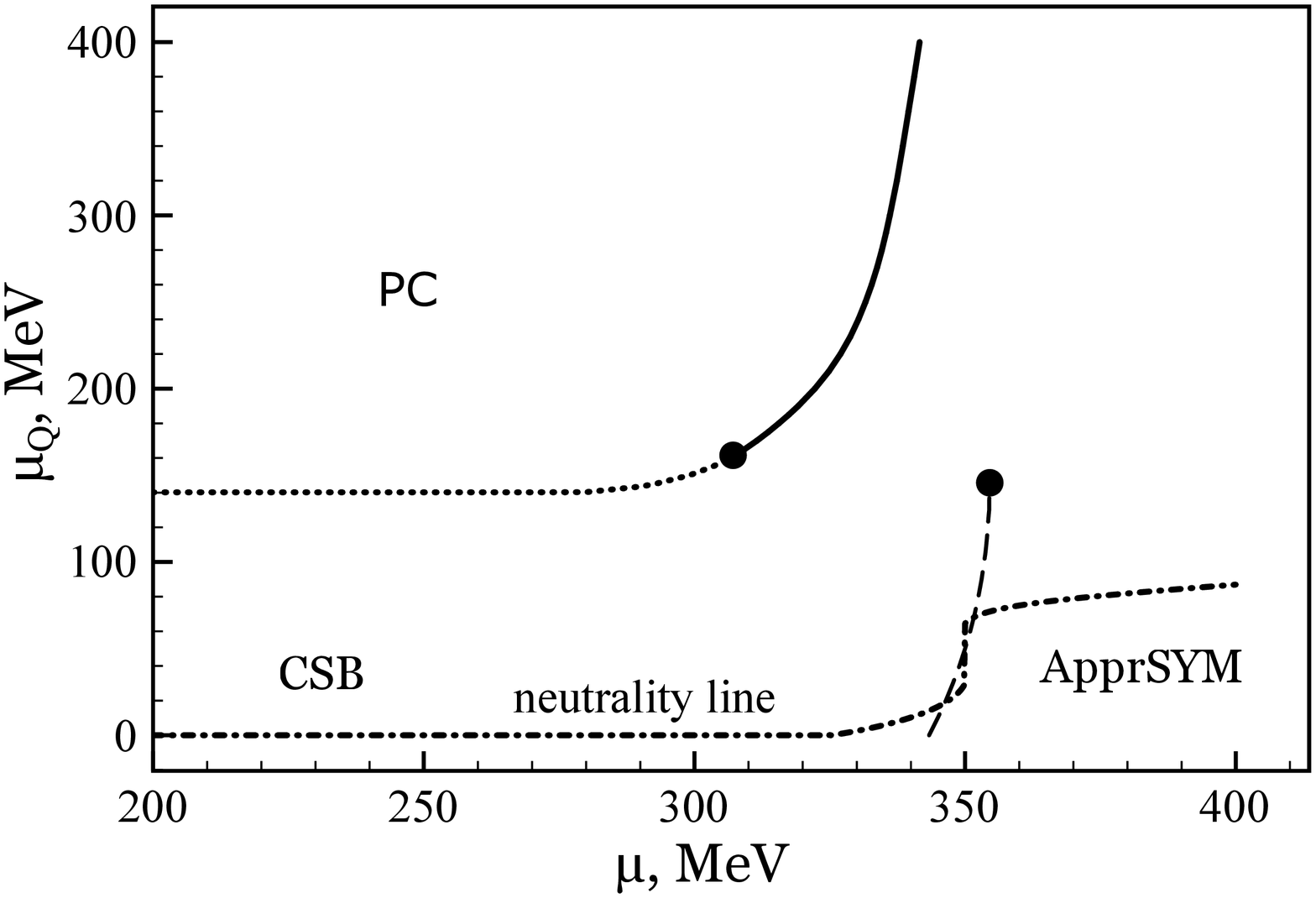}\\
\parbox[t]{0.45\textwidth}{\caption{The ($\mu$,$m_{\pi}$) phase diagram of neutral quark matter. The figure is taken from \cite{Abuki:2008wm}.}}\hfill
\parbox[t]{0.45\textwidth}{\caption{The ($\mu$,$\mu_{Q}$) phase diagram at the physical point, $m=5.5$ MeV. The figure is taken from \cite{Abuki:2008wm}.}}
\end{figure*}

 So let us note that not only PC$_{d}$ phase is absent at the charge neutrality line, but there is no PC at all 
 and electrically neutral quark matter
never meets the regions occupied by the charged PC phase. Thus, one can summarize that at the physical quark mass 
the NJL model consideration of neutral quark matter forbids the charged PC to take place in the ground state.


\section{Conditions promoting the appearance of charged pion condensation in dense quark matter}
In the previous Section 2 (see 2.4) it was shown that charged PC does not take place in neutral quark matter with isospin asymmetry. 
But recently there have been found conditions that can be realized in real physical situations and can lead to the appearance of 
charged PC in dense quark matter \cite{Ebert:2011tt, Gubina:2012wp, Khunjua:2017khh, Ebert:2016hkd, Khunjua:2017mkc, Khunjua:2018sro}. Some of these conditions are elaborated in the framework of the (3+1)-dimensional 
NJL model discussed in the previous section \cite{Khunjua:2017mkc, Khunjua:2018sro}. Other physical constraints that may lead to a generation of the charged
PC phase of dense quark matter are investigated, for simplicity, within the framework of a toy (1+1)-dimensional NJL model 
(NJL$_2$)  \cite{Ebert:2011tt, Gubina:2012wp, Khunjua:2017khh, Ebert:2016hkd}.
So first of all, let us remind the basic properties of the NJL$_2$ model in the presence of baryon $\mu_B\equiv 3\mu$ and isospin 
$\mu_I\equiv 2\nu$ chemical potentials (see, e.g., in Ref. \cite{Ebert:2009ty}). 

\subsection{The NJL$_2$ model and its phase structure}

As it was noted above, the the NJL$_2$ model can be used in order to get quite common qualitative information about 
the properties of real dense quark matter with isospin asymmetry. Formally, its Lagrangian has the same form as the Lagrangian 
(1)-(2) of the NJL model in (3+1)-dimensional spacetime. But $u$ and $d$ quark fields are now (in two dimensions) the 
two-component Dirac spinors and, moreover, the $\gamma^\nu$ ($\nu =0,1$) and $\gamma^5$ matrices are the 2$\times$2 matrices 
in the two-dimensional spinor space,
\begin{equation}
\begin{split}
\gamma^0=\begin{pmatrix}
0&1\\
1&0\\
\end{pmatrix};\quad\;\;\;
\gamma^1=\begin{pmatrix}
0&-1\\
1&0\\
\end{pmatrix};\quad\;\;\;
\gamma^5=\gamma^0\gamma^1=
\begin{pmatrix}
1&0\\
0&{-1}\\
\end{pmatrix}.
\end{split}\label{021}
\end{equation}
It is clear that the NJL$_2$ model is a direct generalization of the (1+1)-dimensional Gross-Neveu model \cite{gn} with a single 
flavor (i. e. single massless color $N_c$-plet) to the case of two flavors of massive quarks. 

To find the TDP of the NJL$_2$ model in the leading order of the large-$N_c$ expansion, one can use the technique of the previous 
section, where we have obtained the corresponding TDP of the (3+1)-dimensional NJL model.
Namely, it is possible to introduce the auxiliary composite bosonic fields $\sigma (x)$ and $\pi_a (x)$ as well as their ground state
expectation values $M-m$ and $\pi_a$, respectively (see in Eqs. (6) and (12), for more details). Moreover, without loss of generality,
it can be assumed that $\pi_1\equiv \Delta\ge 0$ and $\pi_{2,3}=0$. Then the expression $\Omega(M,\Delta)$ for the unrenormalized
TDP of the NJL$_2$ model can be easily shown to have the following form
\begin{eqnarray}
\Omega(M,\Delta)
&=&\frac{(M-m)^2+\Delta^2}{4G}+\mathrm{i}\int\frac{d^2p}{(2\pi)^2}
\ln\Big\{\Big [(p_0+\mu)^2-(E^+_{\Delta})^2\Big ]\Big
[(p_0+\mu)^2-(E^-_{\Delta})^2\Big ]\Big\}, \label{121}
\end{eqnarray}
where $E_\Delta^\pm=\sqrt{(E^\pm)^2+\Delta^2}$, $E^\pm=E\pm\nu$,
$\nu=\mu_I/2$ and $E=\sqrt{p_1^2+M^2}$.
It is possible to perform in Eq. (\ref{121}) the 
integration with respect to $p_{0}$, as in the (3+1)-dimensional case. Then, one can see that the TDP (\ref{121}) is ultraviolet 
divergent, hence to get any physical information, one should first renormalize it (in contrast to the (3+1)-dimensional NJL model,
the NJL$_{2}$ model is renormalizable). For this purpose one can use a special assumption that the bare coupling constant $G$ 
depends on the cutoff parameter $\Lambda$ (where $\Lambda$ restricts the region of integration in the divergent integral 
of (\ref{121}), $|p_1|<\Lambda$). At zero bare quark mass, $m=0$, the renormalization procedure was discussed in the literature, 
see, e.g., in Refs. \cite{Ebert:2011tt,klim,Khunjua:2017khh}. In short, it is evident in this case that cutting off 
the divergent integral in (\ref{121}) and then using
the following dependence of the bare coupling constant on the cutoff parameter, $G\equiv G(\Lambda)$,
\begin{eqnarray}
\frac{1}{2G(\Lambda)}=\frac{2}{\pi}\ln\left
(\frac{2\Lambda}{M_0}\right ), \label{16}
\end{eqnarray}
one has a new free finite renormalization group invariant (does not depend on the cutoff $\Lambda$) massive parameter $M_0$ (and 
it can be shown that it is just a dynamically generated quark mass in vacuum, i.e. at $\mu=0$ and $\nu=0$).
Then in the limit $\Lambda\to\infty$ one can obtain a finite renormalization group invariant expression for the TDP. In particular,
in vacuum, where we denote the TDP as $V_0(M,\Delta)$, it looks like
\begin{eqnarray}
V_0(M,\Delta)
=\frac{M^2+\Delta^2}{2\pi}\left [\ln\left
(\frac{M^2+\Delta^2}{M_0^2}\right )-1\right ]. \label{231}
\end{eqnarray}
It can be shown that the global minimum of the TDP (\ref{231}) is reached at $M=M_{0}$ and $\Delta =0$, and chiral symmetry of 
the model is broken down. In the chiral limit ($m=0$) the phase structure of NJL$_2$ model was studied in detals in Refs. 
\cite{Ebert:2011tt,klim}, and it follows from this considerations that at arbitrary values of $\mu$ and $\nu$ the phase portrait 
of the NJL$_2$ model does not contain the charged PC phase with nonzero density of baryons. The same property is valid for the 
phase structure of the massive NJL$_2$ model, which does not predict the charged PC phenomenon in dense quark matter (see, e.g., in 
Refs. \cite{Ebert:2009ty,Gubina:2012wp}).

Let us note that (1+1)-dimensional NJL model has the same chiral structure as the corresponding (3+1)-dimensional NJL model and 
it possesses a lot of common properties and features with real QCD. For example, QCD and NJL$_2$ models are both renormalizable, 
both have asymptotic freedom, dimensional transmutation and the spontaneous chiral symmetry breaking (in vacuum) 
\cite{wolff,kgk1,barducci,chodos}. 
Moreover, they have similar $\mu_B-T$ phase diagrams.
Hence, NJL$_2$ model can be used as a laboratory for the qualitative simulation of specific features of QCD and, 
since it is renormalizable, it can be done at {\it arbitrary energies} in contrast to effective (3+1)-dimensional NJL model. 

Let us now discuss the breaking of continuous symmetry and the famous {\it no-go} theorem \cite{coleman} that, in general, 
forbids in two-dimensional space-time the spontaneous
breaking of any continuous symmetry.  Currently it is well understood (see, e.g., the discussion in 
\cite{barducci,chodos,thies}) that in the limit  $N_c\to\infty$ ($N_c$ is the number of quark colors) the usual {\it no-go} 
theorem \cite{coleman} does not work. This fact follows directly from the observation that at finite $N_c$ the quantum 
fluctuations, which at finite $N_c$ usually washes away a long range order corresponding to a spontaneous breakdown of symmetry,
are suppressed by the powers of $1/N_c$ factor in the large-$N_c$ limit. Thus, it is possible to simulate the phenomena taking 
place in 
real hadronic (quark) matter, such as spontaneous breaking of continuous chiral symmetry or charged pion condensation 
(spontaneous breaking of the continuous isospin symmetry), in the framework of a comparatively simple
(1+1)-dimensional NJL-type models, although only in the limit of the large $N_c$. 

\subsection{Finite size effect and nontrivial topology}

Now let us return to the factors that can generate the appearance of charged PC in dense quark matter.
Physical phenomena, in general, usually occur 
in a restricted space. For instance, 
the quark gluon matter droplets produced in heavy ion collisions (for example, at RHIC) always has a finite size. A volume of the droplets can range and can be as large as of order of several dozens or even hundred of fm
, while size of the smallest system produced is estimated to be as low as 2 fm \cite{Wang:2018ovx}.
So it is interesting to account for the influence of the finite size effects on ($\mu$,$\mu_{I}$) phase 
diagram and in particular on charged PC and it has been done in \cite{Ebert:2011tt}.
The analysis has been performed in the framework of NJL$_{2}$ model.
In order to study the influence of finite size effects on the phase structure of quark matter, one needs
to assume from the beginning some boundary conditions. In finite
temperature field theory one can introduce temperature by
replacing integral over the temporal direction to a summation of the Matsubara frequencies with the requirement that fermion field satisfies the antiperiodic boundary conditions. But in contrast to the temporal direction (thermal case), where boundary conditions are fixed just by the statistics, there is no
such restriction in the case of spatial directions. So for the system with the finite size there is always important issue concerning the choice of the boundary
conditions for the fermion fields.

In the papers on this subject, both antiperiodic and periodic boundary conditions have been used and neither of them has been 
decisively excluded. Let us emphasize what the most substantial difference between these two boundary conditions is. 
It is actually whether one includes zero-momentum mode contribution or not. The zero-mode contribution is taken into account in the periodic boundary condition, whereas it is excluded in the case of antiperiodic one.
 For bosons the most natural choice of boundary condition is periodic. However, it has been always not clear what boundary 
 conditions one should use for fermions, for example, quarks.
 
 In some cases, the antiperiodic spatial boundary condition is used to keep symmetry between the time and space directions or to get a consistent picture with the dependence of pion mass on the finite size obtained from chiral perturbation theory (ChPT).
 
 Various 
 boundary conditions induce opposite results on vacuum properties of QCD: the antiperiodic spatial boundary condition for 
 quarks induces the chiral symmetry restoration in the small system (analogously to the thermal effects),  while the periodic 
 boundary condition enhances the chiral symmetry breaking in vacuum (similar to the catalysis of chiral symmetry breaking under 
 the influence of magnetic field) \cite{Xu:2019kzy,Inagaki:2019kbc,Inagaki:1997kz}.

There are many discussions of the finite size effects \cite{Xu:2019kzy, MalbouissonMalbouissonPereira, Abreu:2014sfa, Correa:2017vaf, Khanna:2009zz, Khanna:2014qqa, Inagaki:2019kbc, Kiriyama:2005eh, Grunfeld:2017dfu, Bhattacharyya:2014uxa}, and to incorporate it different strategies are employed, for example, replacing momentum integrals by the summations over discrete momentum \cite{Xu:2019kzy, MalbouissonMalbouissonPereira, Abreu:2014sfa, Correa:2017vaf, Khanna:2009zz, Khanna:2014qqa}, introducing a lower momentum cutoff in integrals \cite{Bhattacharyya:2014uxa} or one can use the multiple reflection expansion \cite{Kiriyama:2005eh}. 

\begin{figure*}
\includegraphics[width=0.49\textwidth]{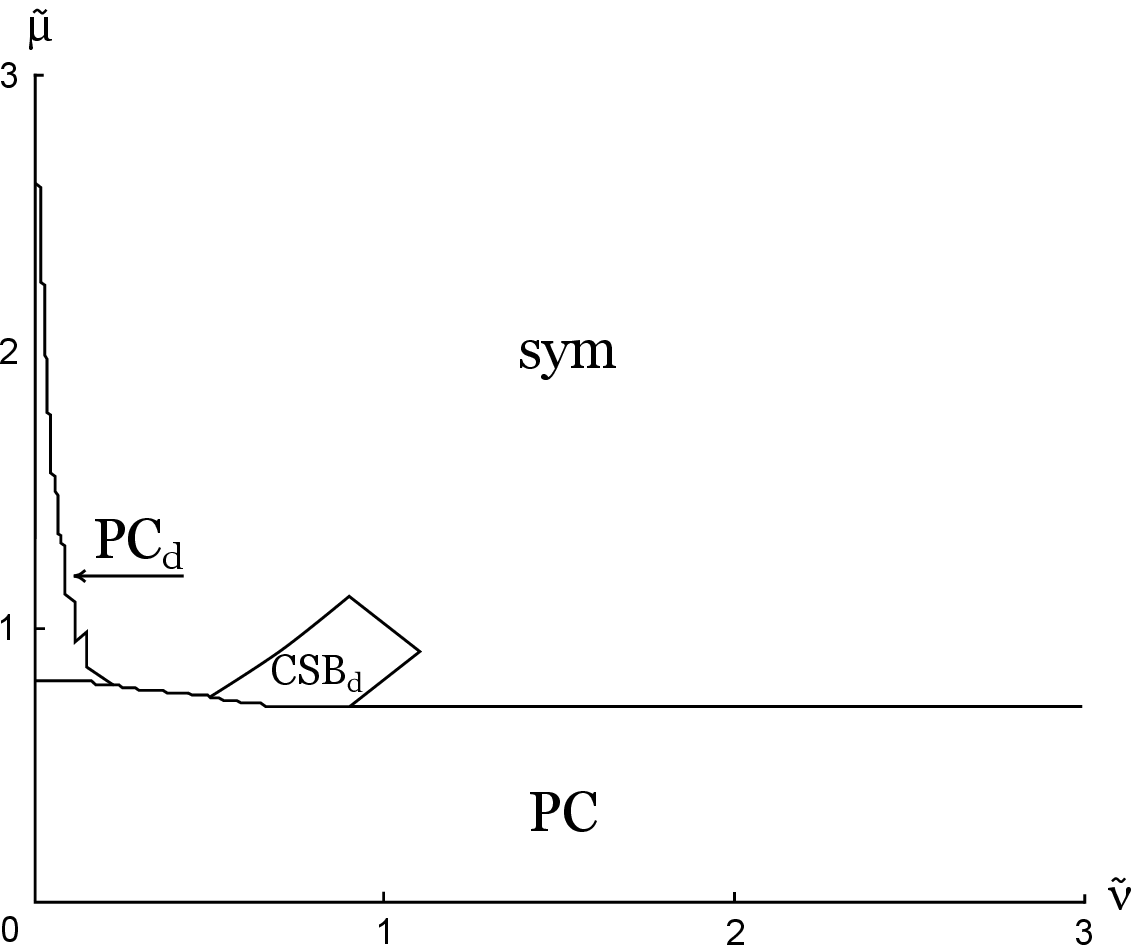}
\hfill
\includegraphics[width=0.49\textwidth]{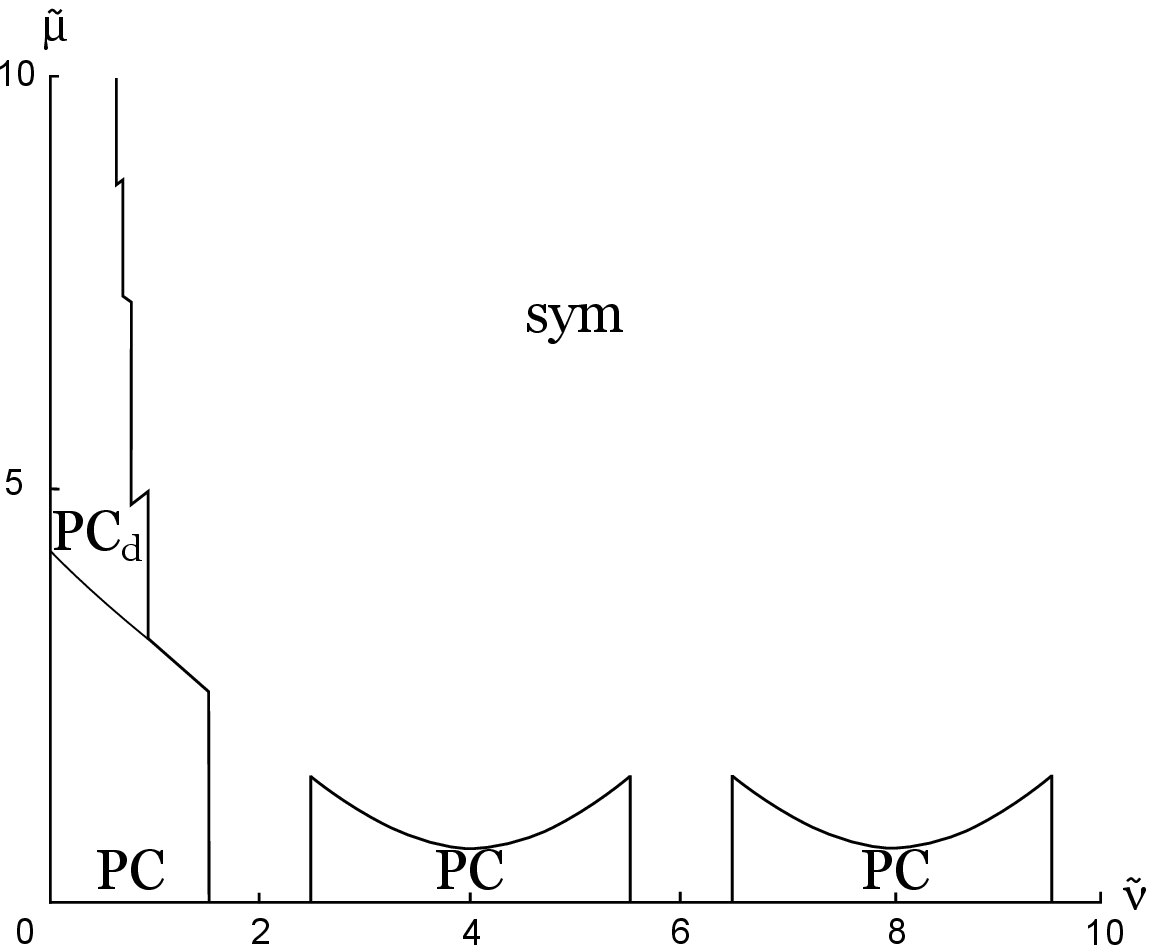}\\
\parbox[t]{0.45\textwidth}{\caption{The $(\tilde\nu,\tilde\mu)$ phase portrait at $\lambda=0.1$ \big(where $\lambda=\frac{\pi}{LM_0}$, 
$\tilde{\mu}=\frac{\mu}{M_0}$ and $\tilde\nu=
\frac{\nu}{M_0}$, see the text \big). The case of periodic boundary conditions.}}\hfill
\parbox[t]{0.45\textwidth}{\caption{The $(\tilde\nu,\tilde\mu)$ phase portrait at $\lambda=2$. The case of periodic boundary conditions.}}
\end{figure*}
The above discussion was devoted to the quark models such as NJL model but finite size effects have been also studied in the framework of different hadronic models such as ChPT, Walecka model, hadron resonance gas model etc.. 
Let us elaborate on that a little bit. In order to get the extrapolation to infinite volume limit for lattice QCD finite size effects have been investigated in the framework of ChPT
(\cite{Colangelo:2004sc} and references therein), 
where the periodic boundary conditions were imposed.
While twisted boundary conditions (similar to the general case of the (\ref{constraint}))  have been used in ChPT in
\cite{Sachrajda:2004mi}.
In terms of hadron resonance gas model, the finite size effects were investigated in \cite{Samanta:2017ohm} employing a lower momentum cutoff, and the multiple reflection expansion formalism has been used to  study influence of  finite size  on  the  chemical  freeze-out  
\cite{Sarkar:2019oyo}.
Walecka model in a finite volume with antiperiodic spatial boundary conditions have been considered in 
\cite{Abreu:2017lxf} and both types of boundary conditions (periodic and antiperiodic) have been investigated in the framework of parity-doublet and Walecka model in \cite{Ishikawa:2018yey}. Other methods not directly involving boundary conditions were used in studies of finite size matter in nuclei \cite{De:1998is, Moretto:2004dm}.

To simulate the influence of finite size effect,  in Ref. \cite{Ebert:2011tt}
the (1+1)-dimensional system with Lagrangian of the form (1)-(2) was placed into a
confined space region of the form $0\le x\le L$ (where $x$ is the
space coordinate). It is well-known that such a constraint is tantamount to the investigation in a space-time with nontrivial topology, for example in our case, one of the space coordinates are compactified. 
It
means that one can consider the model in space-time with $R^1\times S^1$
topology and quark quantum fields should satisfy the
boundary conditions of the following form
\begin{equation}
\begin{split}
q(t,x+L)=e^{i\pi\alpha}q(t,x),
\end{split}\label{constraint}
\end{equation}
where $0\le\alpha\le2$ is the parameter fixing the boundary conditions, for example, $\alpha=0$ corresponds to periodic boundary 
condition and $\alpha=1$ to the antiperiodic one, the parameter $L$ is the length of the circumference $S^1$.

As a result, in order to obtain the TDP $\Omega_{L}(M,\Delta)$ of the NJL$_2$ system with Lagrangian (1)-(2) in the 
confined domain, one should make a standard replacement of the integration over momentum in the expression (\ref{121}) to the 
summation according to the rule:
\begin{equation}
\begin{split}
\int\limits_{-\infty}^{\infty}\frac{dp_1}{2\pi}f(p_1)\rightarrow\frac1L\sum\limits_{n=-\infty}^{\infty}
f(p_{1n}); \quad p_{1n}=\frac{\pi}{L}(2n+\alpha), \quad n=0,\pm1,\pm2,\dots
\end{split}
\end{equation}

Let us briefly discuss the analogy of parameter $\alpha$ with imaginary baryon chemical potential. For temporal dimension the Matsubara frequencies are $p_{0l}=\frac{\pi}{\beta}(2l+1)-i\mu$. Here the antiperiodic boundary conditions have been used. One can note that in the case of imaginary chemical potential the Matsubara frequencies are $p_{0l}=\frac{\pi}{\beta}(2l+1+\text{ Im}(\mu)\beta/\pi)$ and the temporal boundary condition can be transformed into the imaginary chemical potential divided by temperature $\mu/T$. In recent years a great effort has been made in studying the phase diagram at imaginary chemical potential as in lattice QCD \cite{Lombardo:2006yc} as well as in effective models including NJL model (\cite{Kashiwa:2019ihm} and references therein).

Let us briefly discuss the analogy of parameter $\alpha$ with imaginary baryon chemical potential. For temporal dimension the Matsubara frequencies are $p_{0l}=\frac{\pi}{\beta}(2l+1)-i\mu$. Here the antiperiodic boundary conditions have been used. One can note that in the case of imaginary chemical potential the Matsubara frequencies are $p_{0l}=\frac{\pi}{\beta}(2l+1+\text{ Im}(\mu)\beta/\pi)$ and the temporal boundary condition can be transformed into the imaginary chemical potential divided by temperature $\mu/T$. In recent years a great effort has been made in studying the phase diagram at imaginary chemical potential as in lattice QCD \cite{Lombardo:2006yc} as well as in effective models including NJL model (\cite{Kashiwa:2019ihm} and references therein).

Now let us introduce the notations used in \cite{Ebert:2011tt}. Instead of dimensionfull quantities such as $M$, $\Delta$, $\mu$, $\nu$ (see section 3.1) and $L$  it is convenient to use the dimensionless quantities:
\begin{equation}
\begin{split}
\lambda=\frac{\pi}{LM_0},\quad
\tilde{\mu}=\frac{\mu}{M_0},\quad\tilde\nu=
\frac{\nu}{M_0}\equiv\frac{\mu_I}{2M_0}, \quad m=\frac{M}{M_0},\quad
\delta= \frac{\Delta}{M_0}. \label{not}
\end{split}
\end{equation}
So $\lambda$ is just the inverse length of the circumference $S^1$ and shows how large the volume of space, in which our system is confined, is. The case $\lambda=0$ corresponds to the topology $R^1\times R^1$ (infinite space) and the case $\lambda\neq0$ corresponds to $R^1\times S^1$, where the length of compactified dimension is $\frac{\pi}{\alpha M_0}$. One can think in the following way:  the larger $\lambda$ corresponds to smaller volume and stronger finite size effects.

The influence of nontrivial topology on the phase structure of NJL$_{2}$ model has been studied in \cite{Ebert:2011tt}, 
where both periodic and antiperiodic boundary conditions have been studied. Let us present a couple of phase portraits. For example, in
Figures 6 and 7 the ($\nu$,$\mu$) phase diagrams are presented for
various values of $\lambda$ in the case of periodic boundary conditions. And one can see that there is a region of the PC$_{d}$ phase at both plots, it spans around the small values of isospin chemical potential. 
One can also notice that PC$_{d}$ phase is presented in a wide interval of $\lambda$ and it is enlarged if the system becomes 
smaller. In the case of antiperiodic boundary conditions this phase has not been found and it is probably natural, as in this case the 
finite size effect is similar to the effect of the finite temperature on the system (it usually restores all the symmetries and 
leads to the vanishing condensates). But PC$_{d}$ phase has been found in quark matter system of a finite size at least for 
periodic boundary conditions, so probably one can conclude that finite size and non-trivial topology effects might lead to the 
appearance of the charged PC phase with nonzero baryon density at least in the framework of the NJL$_{2}$ model. 
It would be interesting to consider this situation in the framework of more realistic (3+1)-dimensional model approach, 
but unfortunately, to the best of our knowledge, this consideration has not been performed up to now.

Now let us mention another interesting feature of the phase structure characteristic to the finite size systems at zero temperature.
One can see at Figures 6 and 7 that there are small spikes at the border line between PC$_{d}$ and symmetrical phases. At the first glance one can think that they are numerical calculations artifacts. 
But they are actually physical and comes from the fact that 
in the finite systems the momentum spectrum became discrete and 
when one changes external parameters new momentum eigenvalues go into play and this can lead to the non-analyticity of critical lines, condensates and some thermodynamical quantities. Similar oscillations were observed in  
\cite{Ebert:2010eq}.

\subsection{Inhomogeneous pion condensation in dense baryonic matter}

Now let us turn our attention and discuss another factor that can lead to the generation 
of charged PC in dense quark matter, namely, the possibility of inhomogeneity of condensates i. e. dependence of condensates on the spacial space-time dimensions.
The phase structure of dense quark matter with isospin asymmetry has been studied in the framework of NJL$_{2}$ model with 
possibility of inhomogeneous charged pion condensate \cite{Gubina:2012wp}. The consideration was performed at the physical point,
i. e. for physical values of current quark mass.

Let us say a couple of words about why one should consider inhomogeneous condensates. In vacuum, i. e. in empty space without particle 
density and, equivalently, at zero values of the chemical potentials $\mu$ and $\mu_I$, the vacuum expectation values
$\vev{\sigma(x)}$ and $\vev{\pi_a(x)}$ do not depend on space-time coordinates. Although, in dense quark medium or, in other words,
at non-zero values of the chemical potentials $\mu\ne 0$, $\mu_I\ne 0$, the ground state expectation values of bosonic fields 
might assume a nontrivial dependence on spatial coordinate $x$. For example, in the considerations of Ref. \cite{Gubina:2012wp} 
there has been used  the following ansatz 
\begin{eqnarray}
\vev{\sigma(x)}=M-m,~~~\vev{\pi_3(x)}=0,~~~\vev{\pi_1(x)}=\Delta\cos(2bx),~~~ \vev{\pi_2(x)}=\Delta\sin(2bx), \label{6}
\end{eqnarray}
where $M,b$, and $\Delta$ are quantities that do not depend on space-time coordinates.

In \cite{Gubina:2012wp} the phase structure in this approach was explored and it was found that in the inhomogeneous case 
there is charged PC phase with non-zero baryon density at the phase portrait and it is quite extensive. At Figure 8  
the $(\nu,\mu)$ phase portrait of the model NJL$_2$ model is shown. There the phase IPC$_{d}$ are inhomogeneous phases with charged PC and with non-zero baryon density, and one can see that it occupies the large part of the phase diagram. 

So one can conclude that spatial inhomogeneity of charged pion condensate is also a factor promoting the
appearance of PC$_d$ phases at the phase diagram of dense quark matter.
\begin{figure*}
\includegraphics[width=0.59\textwidth]{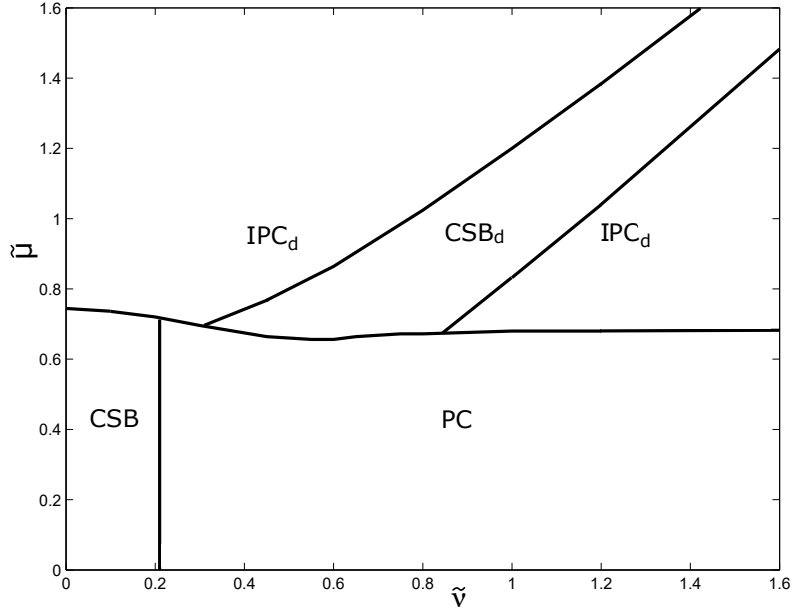}
\centering
\caption{ The $(\tilde{\nu}, \tilde{\mu})$ phase portrait of the model with the possibility of inhomogeneous pion
  condensation. IPC stand for
  inhomogeneous charged PC phases. All other notations are the same. }
\end{figure*}

\subsection{Chiral imbalanced dense quark matter}

Now let us return to the (3+1)-dimensional NJL (1) model and talk about another factor that can generate  charged pion 
condensation in dense quark matter. Besides non-zero temperature and baryon density (nonzero baryon chemical potential), there 
exist additional parameters that may be relevant to the properties of the QCD phase diagram. We have already talked about 
isotopic chemical potential. There is another captivating phenomenon that fell into spotlight quite recently, it is the chiral 
imbalance (different average densities of right- ($n_R$) and left-handed ($n_L$) quarks). This phenomenon is
a remarkable effect originating from highly non-trivial interplay of different features of QCD such as chiral symmetry, chiral (axial) 
anomaly, and possible gluon configuration topology. It is expected to appear in the processes in heavy ion collisions violating  P
and CP on the event-by-event basis \cite{Kharzeev:2007jp} and possibly may lead to interesting phenomena such as chiral magnetic 
effect \cite{Fukushima:2008xe}. Chiral imbalance might be also realized in interiors of compact stars \cite{Khunjua:2018jmn}. 
In addition, chiral media, i. e. media with non-zero chiral imbalance, can
be attained in condensed matter systems such as Dirac and Weyl semimetals \cite{Li:2014bha, Braguta:2019pxt}.
Also, note that phenomena 
regarding a chiral imbalance in the context of QCD are usually studied in the framework of NJL models with a chiral chemical potential \cite{andrianov}.

The chiral imbalance of the system is described by the chiral charge density $n_5 = n_R-n_L$ but instead of it one can use the 
corresponding chiral chemical potential $\mu_{5}$ in the grand canonical ensemble approach. In general, the chiral charge $n_5$ 
is a more relevant quantity, but due to technical reasons when studying the phase diagram, it is easier to work with chemical 
potential $\mu_{5}$, and we will follow this way in this paper. 
Let us also note that the chiral charge $n_5$ is not a strictly conserved quantity, because of non-zero quark condensate as well as quantum chiral anomaly. Therefore, $\mu_{5}$ chemical potential is not a quantity conjugated to a strictly conserved charge. But it is possible to treat $\mu_{5}$ as the chemical potential describing a
system in thermodynamical equilibrium in the state with nonzero value of $n_5$, although, on a larger time scale than the one needed for chirality changing processes to occur.
Moreover, let us talk about other possibility and note that $n_5$ can be introduced for individual quark flavor,
$n_{u5} = n_{uR}- n_{uL}$ and $n_{d5}=n_{dR}-n_{dL}$, and it is evident that
$n_5 = n_{u5} + n_{d5}$. Then it is possible to consider chiral isospin charge $n_{I5}=(n_{u5}-n_{d5})$ as well as the 
corresponding chiral isospin chemical potential $\mu_{I5}$. Unlike the chiral charge $n_5$, the chiral isospin charge $n_{I5}$ 
turns out to be a conserved quantity at least in the massless NJL model. Since gluons interact with different quark 
flavors in exactly the same way, 
it is usually assumed that in QCD matter the chiral charges $n_{u5}$ and $n_{d5}$ are equal and therefore $n_{I5}=0$ 
and $\mu_{I5}=0$. But the chiral isospin chemical potential $\mu_{I5}$ can be generated due to chiral separation effect in dense 
quark matter.

In \cite{Khunjua:2017mkc, Khunjua:2018sro, Khunjua:2018jmn} the phase diagram of dense quark matter both with chiral and 
isospin imbalances has been considered and the influence of chiral $\mu_{5}$ and chiral isospin $\mu_{I5}$ chemical potentials 
on the phase structure and charged PC, in particular, has been studied.
The investigations have been performed in the framework of two flavored (3+1)-dimensional massless NJL model with baryon $\mu_B$, 
isospin $\mu_I$, chiral $\mu_5$ and chiral isospin $\mu_{I5}$ chemical potentials. Its Lagrangian has the form
\begin{eqnarray}
&&  L=L_{NJL}+\bar q\Big [
\frac{\mu_B}{3}\gamma^0+\frac{\mu_I}2 \tau_3\gamma^0+\mu_{5}\gamma^0\gamma^5+\frac{\mu_{I5}}2 \tau_3\gamma^0\gamma^5\Big ]q,
\label{1}
\end{eqnarray}
where $L_{NJL}$ and other notations are presented in Eq. (1). In the following we use the notations 
$\mu\equiv\mu_B/3$, $\nu=\mu_I/2$ and $\nu_{5}=\mu_{I5}/2$.

\begin{figure*}
\includegraphics[width=0.49\textwidth]{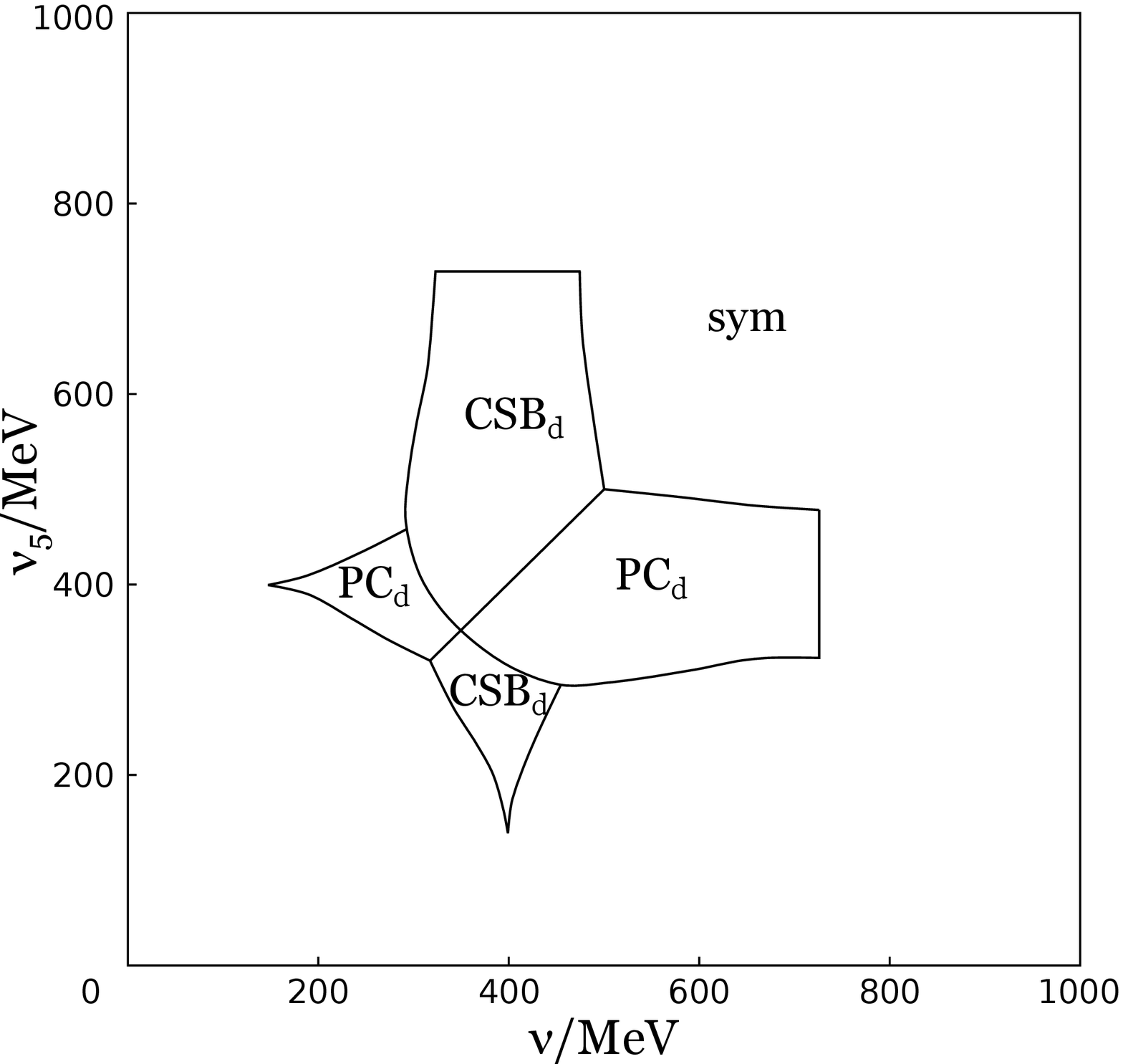}
\hfill
\includegraphics[width=0.49\textwidth]{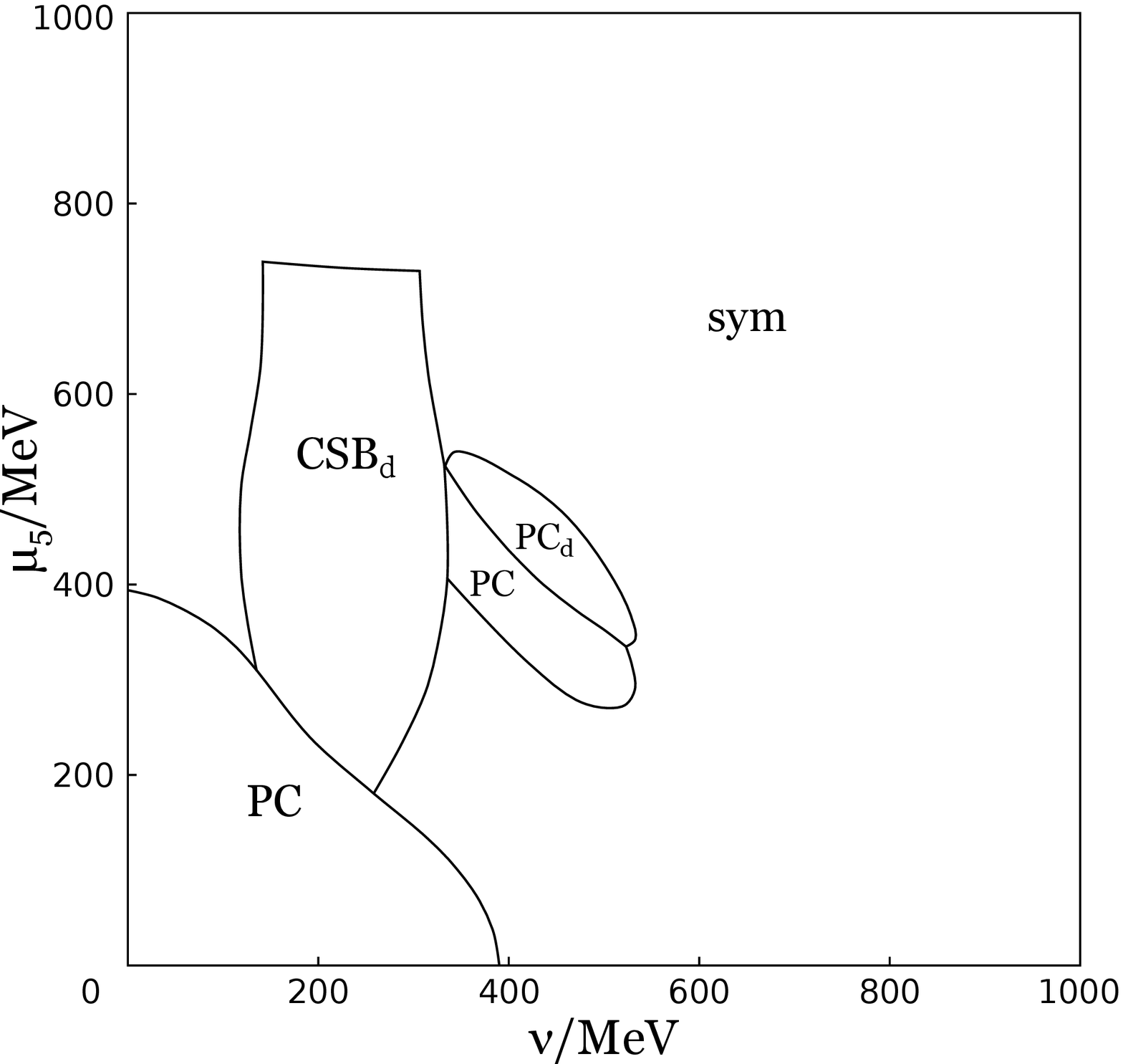}\\
\parbox[t]{0.45\textwidth}{\caption{ $(\nu,\nu_{5})$-phase diagram at $\mu=0.4$ GeV.  The notations are the same as in Figure 1 and 2.}}\hfill
\parbox[t]{0.45\textwidth}{\caption{The $(\nu,\mu_{5})$-phase diagram at $\nu_{5}=0$ and $\mu=0.23$ GeV}}
\end{figure*}

\begin{figure*}
\includegraphics[width=0.49\textwidth]{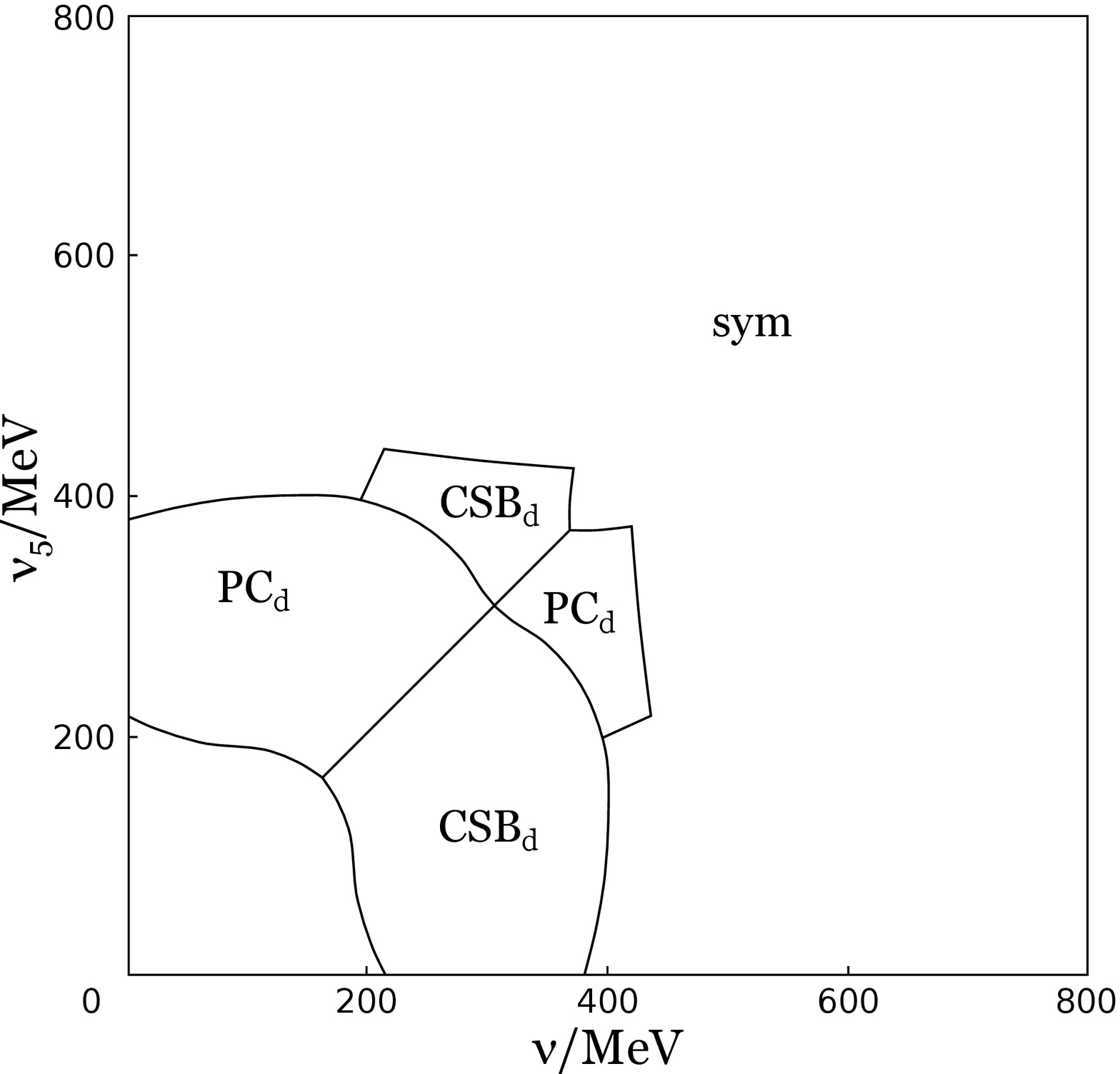}
\hfill
\includegraphics[width=0.49\textwidth]{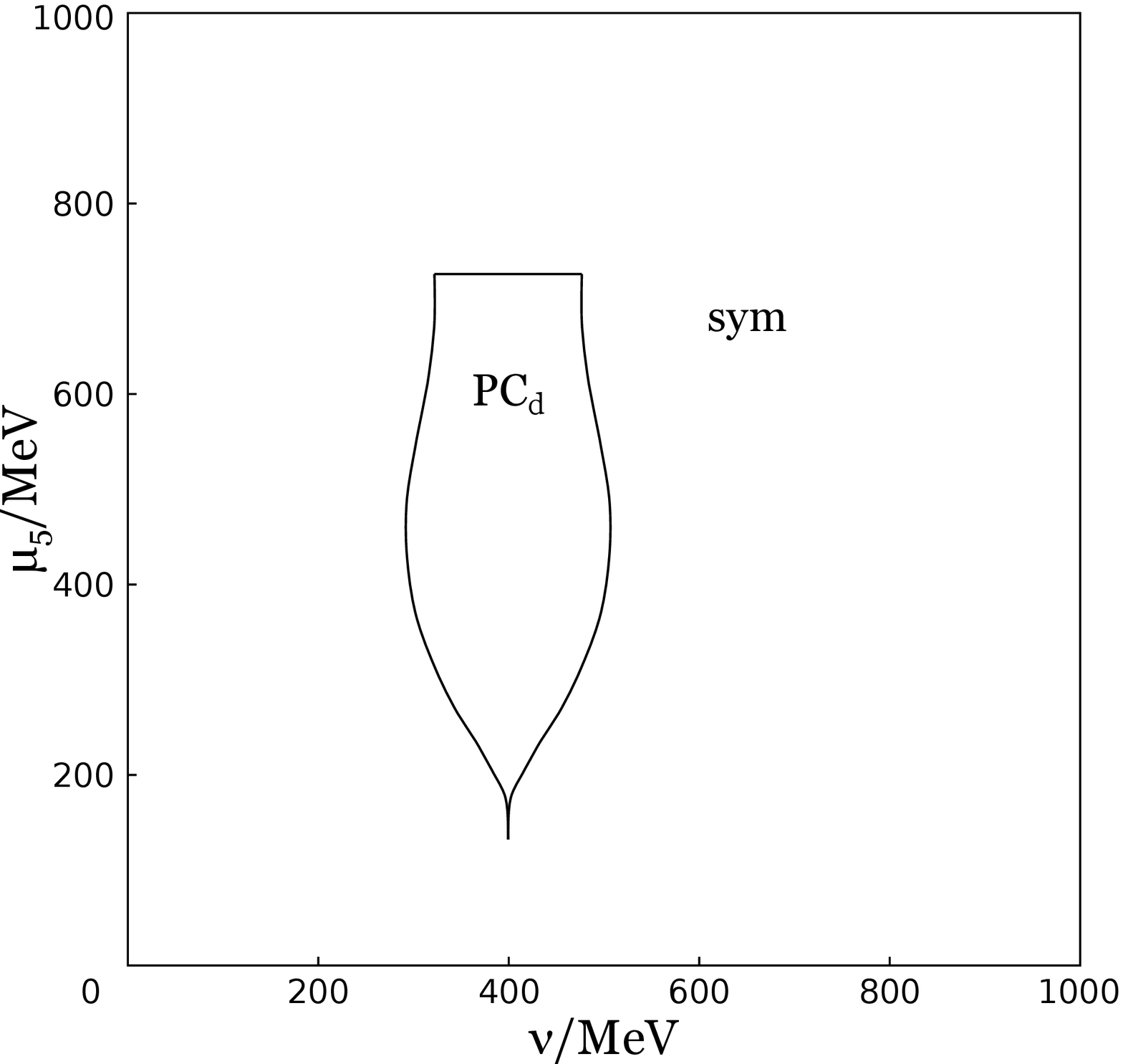}\\
\parbox[t]{0.45\textwidth}{\caption{$(\nu,\nu_{5})$-phase diagram at $\mu=0.43$ GeV and $\mu_{5}=-0.3$ GeV. The notations are the same as in Figure 1 and 2.}}\hfill
\parbox[t]{0.45\textwidth}{\caption{The $(\nu_{5},\mu_{5})$-phase diagram at $\nu=0$ and $\mu=0.4$ GeV.}}
\end{figure*}

The phase diagram of NJL model with baryon $\mu_{B}$, isospin $\mu_{I}$, and chiral isospin $\mu_{I5}$ chemical potentials (at $\mu_5=0$)
has been considered in \cite{Khunjua:2017mkc} and the $(\nu,\nu_{5})$ phase portrait of the model is shown at Figure 9. 
It is clear from this figure that chiral imbalance (chiral isospin chemical potential $\mu_{I5}$) leads to the generation of 
charged PC in 
dense quark matter with isotopic asymmetry. But let us note that for the PC$_{d}$ phase to be generated the matter should have non-zero isospin 
imbalance (isospin chemical potential $\mu_{I}$). One can also observe that in contrast to the first prediction of PC$_{d}$ phase in dense quark matter with isotopic asymmetry (see Sections 2.2, 2.4 and Figures 1, 3), the generation is quite extensive and the region of PC$_{d}$ phase is quite large (not 
marginal in any way).

Then in \cite{Khunjua:2018sro} the influence of chiral chemical potential $\mu_{5}$  was studied in addition, and it was shown 
that chiral chemical potential $\mu_{5}$ generates charged PC phase in dense baryonic (quark) matter as well. It is easily seen from Figure 10,
where the $(\nu,\mu_{5})$ phase 
portrait of the NJL model is presented at $\nu_5=0$. But this generation is not that vast and occurs at not so large baryon 
densities. The case of the joint effect of two chiral chemical potentials, $\mu_{5}\ne 0$ and $\nu_{5}\ne 0$,  on the phase 
structure of the NJL model was also considered in this work.
The $(\nu,\nu_{5})$-phase diagram (at $\mu_{5}\ne 0$) is depicted at Figure 11 and one can see that at non-zero values of 
$\mu_{5}$ there is a large patch of PC$_{d}$ phase. Moreover, this phase can be generated even at zero values of isospin 
chemical potential $\mu_{I}$. The fact that there is no need for isospin imbalance in order to generate the PC$_{d}$ phase is clearly 
presented at Figure 12, where the $(\nu_{5},\mu_{5})$-phase diagram is depicted at zero value of isospin chemical potential, 
$\nu=0$, and it exhibits a large region containing PC$_{d}$ phase. So one can conclude that chiral and chiral isospin chemical 
potentials together generate PC$_{d}$ phase very efficiently.  
Summarizing, one can say that chiral imbalance is a factor that leads to the generation of charged PC in
dense quark matter.

\section{Summary and Conclusions}
In this short review we tried to give an outline of investigations of charged PC in dense baryonic (quark) matter 
in the framework of effective (3+1)-dimensional NJL model and toy QCD motivated (1+1)-dimensional NJL model (NJL$_2$ model).

The possibility of charged PC phase with non-zero baryon density was shown for the first time in the framework of the massless NJL model in Ref.
\cite{Ebert:2005cs} and then in Ref. \cite{Ebert:2005wr} it was found that this phase persists even in 
electrically neutral dense baryonic (quark) matter. But later it was shown in Refs. 
\cite{Abuki:2008wm, Anglani:2010ye, Abuki:2008tx, Andersen:2007qv, Abuki:2008nm} that charged PC condensation in neutral dense 
quark matter is 
enormously frail with respect to explicit chiral symmetry breaking effect (non-zero current quark mass), and it was shown, in particular, that it is 
forbidden for the physical values of current quark masses.

Nevertheless,  
recently there have been found factors and conditions that can be realized in real physical systems and can promote the 
appearance of charged PC phenomenon in dense baryonic (quark) matter. Namely, in the paper \cite{Ebert:2011tt} it was 
shown that if one includes into consideration the fact that system can have finite size, then charged PC phase with non-zero
baryon density can be realized in the system. It was also revealed in \cite{Gubina:2012wp} that the possibility of spatially
inhomogeneous charged PC condensate allows this phase to appear as well. And more recently 
\cite{Khunjua:2017khh, Khunjua:2018sro, Khunjua:2017mkc, Ebert:2016hkd} it was found that there is another interesting 
factor that can induce a charged PC phase in dense baryonic matter, it is a chiral imbalance of the system 
(nonzero difference between densities of left-handed and right-handed quarks).

The finite size effect and spatially inhomogeneous charged PC have been considered in Refs. \cite{Ebert:2011tt, Gubina:2012wp}
in the framework of a toy 
QCD motivated NJL$_{2}$ model, and the influence of chiral imbalance has been investigated both in the framework of 
NJL$_{2}$ model \cite{Khunjua:2017khh,  Ebert:2016hkd} and in more realistic (3+1)-dimensional NJL model 
\cite{Khunjua:2018sro, Khunjua:2017mkc}. These models allow one to consider the region of non-zero baryon densities of 
QCD phase diagram, which up to now cannot be studied in lattice QCD (simulations are almost impossible due to the infamous 
sign problem).  For simplicity, the whole review is devoted to the consideration of the case of zero temperature only. However,
our recent investigation \cite{Khunjua:2018jmn}, made in the framework of the NJL$_4$ model with non-zero bare quark masses,
shows that in chirally asymmetric dense quark matter charged PC phenomenon can be realized even at 
sufficiently high temperatures.

\vspace{15pt}

Let us once more recapitulate the conditions that can lead to the generation of charged PC phenomenon in dense baryonic matter:

\vspace{7pt}

1) finite size effects and non-trivial topology,

\vspace{3pt}

2) possibility of inhomogeneous charged pion condensates,

\vspace{3pt}

3) chiral imbalance.

\vspace{7pt}

These results can be interesting in the context of heavy ion collision experiments 
such as NICA and FAIR, where it is expected to get high baryon densities. It is of interest also in the neutron star physics, since quark matter might be produced in their cores, where very high baryon and isospin densities are attained.

\section{Acknowledgements}
We would like to thank Sergei D. Odintsov for information about this special issue. And we are especially gratefull to Tomohiro Inagaki for the opportunity to contribute.

R.N.Z. is grateful for support of the Foundation for the Advancement of Theoretical Physics and Mathematics
BASIS grant and Russian Science Foundation under the grant No 19-72-00077


\end{document}